\documentclass[fleqn,usenatbib]{mnras}

\usepackage{newtxtext,newtxmath}

\usepackage[T1]{fontenc}

\DeclareRobustCommand{\VAN}[3]{#2}
\let\VANthebibliography\thebibliography
\def\thebibliography{\DeclareRobustCommand{\VAN}[3]{##3}\VANthebibliography}

\usepackage{graphicx}	
\usepackage{amsmath}	
\usepackage{lipsum}  
\usepackage{gensymb}
\usepackage[normalem]{ulem}

\title[Synthetic spatially resolved spectroscopy]{Spatially resolved mock observations of stellar kinematics: full radiative transfer treatment of simulated galaxies}

\author[D. Barrientos Acevedo et al.]{
Daniela Barrientos Acevedo,$^{1,2}$\thanks{E-mail: daniela.barrientosacevedo@ugent.be}
Arjen van der Wel,$^{1}$
Maarten Baes,$^{1}$
Robert J.J. Grand,$^{3,4,5}$
\newauthor
~Anand Utsav Kapoor,$^{1}$
Peter Camps,$^{1}$
Anna de Graaff,$^{6}$
Caroline M. S. Straatman,$^{1}$
and Rachel Bezanson$^{7}$
\\
$^{1}$ Sterrenkundig Observatorium, Universiteit Gent, Krijgslaan 281 S9, B-9000 Gent, Belgium\\
$^{2}$ European Southern Observatory, Karl-Schwarzschild-Str. 2, D-85748 Garching, Germany\\
$^{3}$Instituto de Astrof\'isica de Canarias, Calle Vía L\'actea s/n, E-38205 La Laguna, Tenerife, Spain\\
$^{4}$Departamento de Astrof\'isica, Universidad de La Laguna, Av. del Astrof\'isico Francisco S\'anchez s/n, E-38206, La Laguna, Tenerife, Spain\\
$^{5}$Astrophysics Research Institute, Liverpool John Moores University, 146 Brownlow Hill, Liverpool, L3 5RF, UK\\
$^{6}$Max-Planck-Institut f\"ur Astronomie, K\"onigstuhl 17, D-69117, Heidelberg, Germany\\
$^{7}$University of Pittsburgh, Department of Physics and Astronomy, 100 Allen Hall, 3941 O’Hara St., Pittsburgh PA 15260, USA
}

\date{Accepted XXX. Received YYY; in original form ZZZ}

\pubyear{2023}

\begin{document}
\label{firstpage}
\pagerange{\pageref{firstpage}--\pageref{lastpage}}
\maketitle

\begin{abstract}
We present a framework to build realistic mock spectroscopic observations for state-of-the-art hydrodynamical simulations, using high spectral resolution stellar population models and full radiative transfer treatment with \texttt{SKIRT}. As a first application we generate stellar continuum mock observations for the Auriga cosmological zoom simulations emulating integral-field observations from the SAMI Galaxy survey. We perform spectral fitting on our synthetic cubes and compute the resulting rotation velocity ($V_{\rm{rot}}$) and velocity dispersion within 1$R_{\text{e}}$ ($\sigma_{\text{e}}$) for a sub-set of the Auriga sample. We find that the kinematics produced by Auriga are in good agreement with the observations from the SAMI Galaxy survey after taking into account the effects of dust and the systematics produced by the observation limitations. We also explore the effects of seeing convolution, inclination, and attenuation on the line-of-sight velocity distribution. For highly inclined galaxies, these effects can lead to an artificial decrease in the measured $V/\sigma$ by nearly a factor two (after inclination correction). We also demonstrate the utility of our method for high-redshift galaxies by emulating spatially resolved continuum spectra from the LEGA-C survey and, looking forward, E-ELT HARMONI. Our framework represents a crucial link between the ground truth for stellar populations and kinematics in simulations and the observed stellar continuum observations at low and high redshift.

\end{abstract}




\section{Introduction}\label{sec:intro}

Spatially resolved spectroscopy has become a fundamental tool for understanding the formation and evolution of galaxies. The continuum and its absorption features allow us to probe the stellar population properties, such as ages and metallicities, but also their morphologies through their kinematics. By looking at the spatially resolved kinematics of galaxies, we can learn about their assembly histories, as the different mechanisms that drive the growth in galaxies leave an imprint on their stellar and gas distributions.

Early works using integral field spectroscopy (IFS) surveys to observe the stellar kinematics, such as~\citet{Emsellem2004TheGalaxies} using the SAURON~\citep{Bacon2001TheSpectrograph} instrument and later ATLAS$^{\rm{3D}}$~\citep{Cappellari2011TheCriteria}, started to uncover formerly hidden structures in the two-dimensional stellar distribution of some systems. Systems previously thought of as smooth and structureless by their light distributions, such as Early-type galaxies (ETG), when looking at the kinematics showed a rotational component~\citep{Emsellem2007TheGalaxies,Cappellari2007TheKinematics}. Several other irregular and complex structures were also revealed, increasing the number of kinematic classes and adding more complexity to the picture of galaxy evolution.

More recently a new generation of IFS surveys, such as CALIFA (Calar Alto Legacy Integral-Field Area,~\citet{Sanchez2016CALIFASurvey}), MaNGA (Mapping Nearby Galaxies at Apache Point, ~\citet{Bundy2015OverviewObservatory}) and the SAMI Galaxy Survey (Sydney-AOO Multi-object Integral field spectrograph, ~\citet{Bryant2015TheSelection}), are finding exciting new results associating these structures to other physical properties. Links between the morphological, kinematical and star-formation state of galaxies have been observed~\citep{Cortese2016TheMorphology, vandeSande2017TheKinematics, Barrera-Ballesteros2015TracingStage, Graham2018SDSS-IVProperties, Fraser-McKelvie2021ASequence} helping to disentangle the connections between mass growth and the internal and external drivers of it.  

However these observations are affected by the angular resolution limitations set by the atmospheric conditions. Due to beam smearing the kinematic distribution gets blended leading to an underestimation and overestimation of the line-of-sight velocity and velocity dispersion respectively. As the inner profile is generally steeper than the outskirts, the difference in apertures in observations can lead to biases in the observed trends~\citep{Graham2018SDSS-IVProperties,vandeSande2017TheSurveys,Harborne2020RecoveringData}. These issues become more dire at high redshift as a consequence of the smaller apparent sizes of the targets.

Most studies of the stellar continuum target present-day galaxies (at $z \lesssim 0.1$) as obtaining high signal-to-noise (S/N) observations of this kind at high $z$ requires long integration times. Nonetheless, the Large Early Galaxy Census survey (LEGA-C,~\citet{vanderWel2016THEGyr}) exploited the capabilities of the Very Large Telescope's (VLT) VIsible Multi-Object Spectrograph (VIMOS) to observe the stellar continuum of $\sim$ 4000 galaxies at $z\sim1$ in the COSMOS field for the first time. Using these data,~\citet{Bezanson2018Galaxies} studied the dynamical state of a sample of quiescent galaxies and compared them to their local counterparts from the CALIFA survey, finding evolution in the rotational support. However, the distances to these galaxies and the ground based nature of the observations result in angular resolutions similar to the sizes of the sources. This poses a challenge for the measurement of kinematics, especially in the central regions of the galaxies.

The recently launched James Webb Space Telescope (JWST) and the potential future large ground based telescopes such as ESO's Extremely Large Telescope (ELT), the Giant Magellan Telescope (GMT) in Chile, and Thirty Meter Telescope (TMT) in Hawaii, will allow for the exploration of the galaxy populations at an even earlier cosmic epoch ($z>2$) and unprecedented spatial resolutions. In order to account for the effects of seeing and inclination beyond the present-day Universe we must rely on dynamical models~\citep{vanHoudt2021StellarLEGA-C}, but for galaxies with high star-formation rates and large dust columns, as well as galaxies with irregular morphologies, the commonly made underlying assumptions in such models (e.g., mass follows light) are no longer valid. 

Cosmological hydrodynamical simulations then play a key role in interpreting the observations of galaxies across cosmic time. These simulations come in two flavors. On one hand, large-volume simulations produce large samples of galaxies at the cost of a limited spatial resolution, while zoom-in simulations focus on recreating individual galaxies and the smaller scale processes~\citep{Vogelsberger2020CosmologicalFormation}.

A new generation of cosmological simulations, both of the large-volume and zoom-in kind, such as the Auriga Project~\citep{Grand2017TheTime}, FIRE-2~\citep{Hopkins2018FIRE-2Formation} (zoom-in), IllustrisTNG~\citep{Pillepich2018FirstGalaxies,Nelson2018FirstBimodality, Naiman2018FirstEuropium}, \texttt{ARTEMIS}~\citep{Font2020TheGalaxies} and \texttt{NewHorizon}~\citep{Dubois2021IntroducingTime} (large-volume), generate galaxies with resolutions that permit the computation of spatially resolved quantities while also providing large enough samples to allow for statistical comparisons with the observed surveys. 

Although the spatial resolution has increased substantially, these simulations still rely on sub-grid recipes to implement many of the feedback mechanisms that act on much smaller scales but that control the star-formation process, illustrating the need for observations to test and constrain the models. In order for this feedback loop to function we need a tool to compare both (that creates realistic mock observations of the stellar continuum spectrum) accounting for limitations in spatial and spectral resolution, projection effects and attenuation (seen/found/expected in observations).

There are many methods to create mock observations, from simple, flexible and computationally cheap to complex and expensive. Among the simpler methods, is the projection and binning of particle data to directly obtain line-of-sight velocity distribution (LOSVD) mass-weighted moment maps. Complexity can be added to this method by assigning single stellar populations to each particle and light-weighting the resulting maps or using a convolution kernel to mimic seeing effects. This approach was used by~\citet{Tescari2018TheSimulations} to generate synthetic gas rotation velocity and velocity dispersion maps of the EAGLE galaxies and make comparisons with the SAMI Galaxy Survey data. Another example of a similar procedure is presented in~\citet{Harborne2020SimSpin-ConstructingCubes}, their software, \texttt{SimSpin}, was used to generate mock observations of various types of galaxies at multiple spatial resolutions to compute corrections for the offsets produced in $\lambda_{\text{R}}$ and $V/\sigma$ measurements due to seeing limitations. Recently, \citet{Bottrell2022RealisticRealSim-IFS} introduced \texttt{RealSim-IFS}, a tool that uses this method to generate IFS-like observations from simulated data and applied it to 893 galaxies from the TNG50 simulation to generate MaNGA-like kinematics.

More complex methods involve the generation of IFS-like spectra from which kinematic properties can be retrieved using the same tools and models applied to observed data.~\citet{Guidi2018TheSimulations} introduced a method to generate mock spectra and applied it to 3 galaxies emulating the CALIFA survey setup, including stellar and nebular emission with kinematic broadening. \citet{Ibarra-Medel2019OpticalMethod} produced MaNGA-like spectra for two simulated galaxies and explored the effects that inclination, resolution, signal-to-noise ratio and attenuation have on the recovered stellar mass and age radial profiles. More recently,~\citet{Nanni2022IMaNGA:Cubes} presented \texttt{iMaNGA}, a method to generate mock MaNGA spectroscopic observations including all the systematics found in the survey data. Using galaxies from the TNG50 simulation they generated a catalogue of 1000 MaNGA mock spectra from which they measured ages, metallicities and kinematics using spectral modelling~\citep{Nanni2023IMaNGA:Catalogue}, reporting consistency within 1-$\sigma$ between the intrinsic galaxy properties and those derived through their method. Using the same simulation,~\citet{Sarmiento2023MaNGIA:Analysis} present MaNGIA, a catalogue of mock observations matching the MaNGA survey instrumentals and target selection, designed to study the star formation history, age, metallicity, mass and kinematics of these galaxies.

Even though many of the mentioned studies strive to replicate the observational conditions of their targeted surveys, most of them tend to overlook the impact of interstellar dust on the radiation emitted by the stars~\citep{Bottrell2022RealisticRealSim-IFS} or often apply simple, general recipes to reproduce it~\citep{Ibarra-Medel2019OpticalMethod, Sarmiento2023MaNGIA:Analysis}. Stellar light can be absorbed or scattered by dust grains in the interstellar medium and these events affect as much as 30-50\% of the light emitted in the Universe~\citep{Viaene2016TheSurvey}. This processing can result in the distortion of optical images and the structural and morphological parameters derived from them (e.g., \citet{Mollenhoff2006ModellingWavelengths, Gadotti2010RadiativeParameters, Pastrav2013TheBulges, Pastrav2013TheBulgesb, deGraaff2022ObservedObservations}). In~\citet{Baes2003RadiativeGalaxies}, the authors showed how the introduction of dust modified the shape of the rotation curves of disk galaxy models as well as their dispersion profiles. These discrepancies increase with increasing inclination of the disk and optical depth, so it is expected that at high redshift ($z\gtrsim1$), where star forming galaxies are gas-rich and dusty, these effects would be exacerbated. In addition, due to limited spatial resolution, it will be impossible to reconstruct the intrinsic stellar mass and dust distribution from the observations alone. Out of the studies listed, only~\citet{Guidi2018TheSimulations} and~\citet{Nanni2022IMaNGA:Cubes} use some form of radiative transfer in the generation of their mocks, but their methodologies consider attenuation effects on the resulting fluxes and disregard the effect the interaction between the photons and dust particles has on the Doppler shift that would lead to changes in the observed kinematics. \citet{Nevin2021AccurateKinematics}, on the other hand, use the code \texttt{SUNRISE}~\citep{Jonsson2006SUNRISE:Geometries, Jonsson2010High-resolutionDust} to incorporate radiative transfer in their mock MaNGA spectroscopy. Despite not being the primary focus of their study, they note how dust attenuation and scattering perturb the 2D kinematic distribution of their simulated galaxies, highlighting the need for a proper dust treatment in the production of mock spectroscopy.

We address this issue by creating a framework for producing synthetic spatially resolved spectra from cosmological simulations using radiative transfer to simulate the effects of the interestellar medium (ISM). In this work we make use of the 3D Monte Carlo radiative transfer code \texttt{SKIRT}~\citep{Camps2020SKIRTGrains}, that in its latest version allows for the inclusion of velocity information in the radiative transfer process and therefore produces spectra including these signatures. The code has been used previously to generate synthetic observations of hydrodynamical simulations, including studies based on integrated fluxes~\citep{Camps2016Far-infraredSimulations,Camps2018DataProject,Trayford2017OpticalSKIRT,Trcka2020ReproducingSample,Baes2020InfraredSimulation,Katsianis2020TheSimulations,Trcka2022UVGalaxies,Vijayan2022FirstDawn} to study luminosity functions and other physical properties derived from the multiwavelenght photometry and also spatially resolved imaging~\citep{Rodriguez-Gomez2019TheObservations,Bignone2020Non-parametricSimulation,Kapoor2021High-resolutionProject,Whitney2021Galaxy3,Camps2022High-resolutionProject, Guzman-Ortega2022MorphologicalGalaxies}, to look at the morphology and sizes of the simulated galaxies. 

We use our methodology to produce synthetic observations imitating the specifications of the SAMI Galaxy survey using as input the simulated galaxies from the Auriga Project. We illustrate the capabilities of our framework by describing discrepancies in the recovered absorption line features and kinematics when the effects of limited seeing conditions, spatial resolutions and attenuation are accounted for. We use these derived kinematics to compare with observed scaling relations obtained using the SAMI Galaxy Survey data and explore the sources of bias in the relations. We also test our method at higher $z$, by mocking the LEGA-C survey at $z\sim 1$, and future observations with the E-ELT and assess the dominant sources of bias in the recovered rotational support at each of these observational set-ups. 

In Section~\ref{sec:auriga}, we give a general description of the Auriga Project and the properties of the galaxies used in this work. In Section~\ref{sec:skirt}, we give some background on the radiative transfer code \texttt{SKIRT} and the specific configuration and parameters used to produce our synthetic spectra. Further information about the addition of instrumental effects to reproduce the observations in the SAMI Galaxy Survey is provided in Section~\ref{sec:final-steps}. The analysis of the mock spectroscopy's kinematics and comparison with the observational sample it emulates is shown in Section~\ref{sec:sami-compare} and in Section~\ref{sec:dust-vs-instrumental} we further explore the sources of bias in the global kinematics produced by our methodology. Section~\ref{sec:high-z} contains the higher $z$ test for LEGA-C and HARMONI spectroscopy and its analysis and lastly, in Section~\ref{sec:summary} we summarise our findings and give an overview of the capabilities of our method and possible applications.


\section{The AURIGA Simulation}\label{sec:auriga}

As the input model for our synthetic observations we used the galaxies from the Auriga Project~\citep[][]{Grand2017TheTime}, a suite of 30 cosmological magneto-hydrodynamical zoom simulations of Milky Way-sized galaxies, simulated in a cosmological environment. The simulations were performed using the moving-mesh code \texttt{AREPO}~\citep[][]{Springel2010EMesh} and following a $\Lambda$ Cold Dark Mater ($\Lambda$CDM) cosmology with parameters $h = 0.693$, $\Omega_m=0.307$, $\Omega_b=0.048$ and $\Omega_{\Lambda}=0.693$ \citep{PlanckCollaboration2013PlanckParameters}. The simulations were run at different resolution levels, with level 4 being the ``standard'' version and the one we used for this work. At this level the typical mass resolutions for the high-resolution dark matter and baryonic particles are~$\sim$3$\times10^5 \rm{M_{\odot}}$ and $\sim$5$\times10^4 \rm{M_{\odot}}$ respectively. The gravitational softening length for the dark matter and star particles evolves with the scalefactor, reaching a maximum value of 369 pc at $z=1$ after which it remains constant. For gas cells, the softening length is scaled by the mean radius of the cell, and it can range from 369 pc to 1.85 kpc, from $z=1$ to $z=0$. 
The galaxy formation model implemented in these simulations follows the evolution of the different components in the halo, including prescriptions for the ISM evolution and subsequent star formation which occurs stochastically and following a~\cite{Chabrier2003GalacticFunction} initial mass function (IMF). It also includes stellar and chemical evolution, stellar and AGN feedback and the seeding and evolution of magnetic fields. 

This comprehensive physical model coupled with the high resolution of the baryonic particles makes Auriga a very well suited sample of simulated galaxies to explore the resolved physical properties, such as the kinematics. 

\begin{figure*}
\begin{center}
    \includegraphics[width=\textwidth]{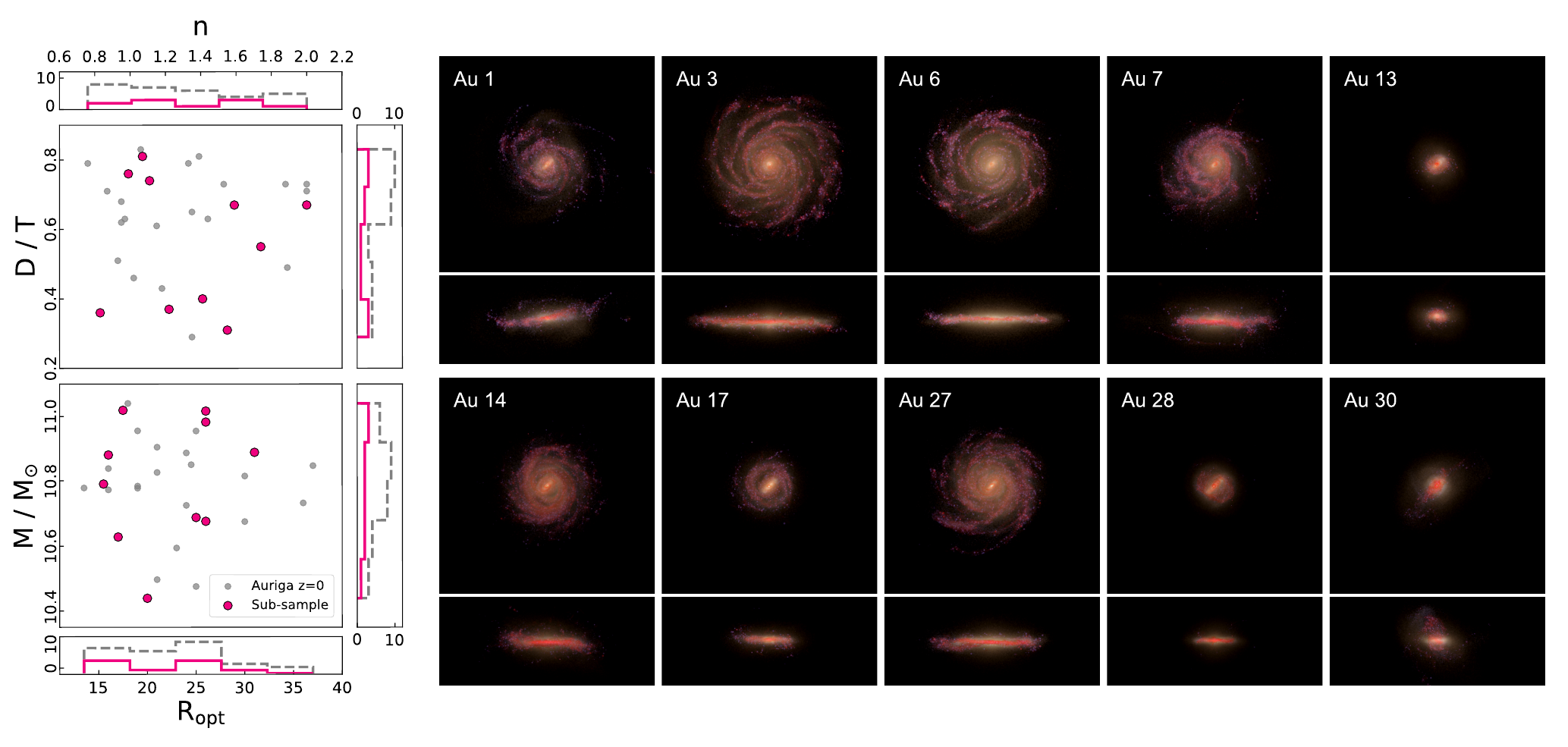}
    \caption{On the left: Physical properties of the galaxies in the Auriga sample at $z=0$ (grey). Stellar masses, optical radii, disk-to-total mass ratios and the Sersic indexes of the bulges where taken from Table 1 in~\citet{Grand2017TheTime}. Pink dots indicate the sub-sample of galaxies used to generate the $z=0$ SAMI-like synthetic observations used in this study. On the right: Face-on and edge-on synthetic RGB image cutouts of the 10 galaxies selected at $z=0$ for this study. The post-processed images were generated by~\citet{Kapoor2021High-resolutionProject} and use, a combination of SDSS-u, -g, -r and -z, for the optical channel, GALEX-FUV, for blue and SPIRE-250 $\mu$m, for the red channel.}\label{fig:subsample}
\end{center}
\end{figure*}

Since the aim of the Auriga project was to simulate Milky Way-like galaxies, the sample at $z=0$ has the properties of typical late-type discs, with masses ranging between $10^{10}$ and $10^{11}\rm{M_{\sun}}$ and mostly sub-dominant bulge components, as illustrated in Figure~\ref{fig:subsample} where only $\sim17\%$ (5) of the galaxies show disk-to-total mass ratio values of $D/T <= 0.4$.

\begin{figure*}
\begin{center}
     \includegraphics[width = \textwidth]{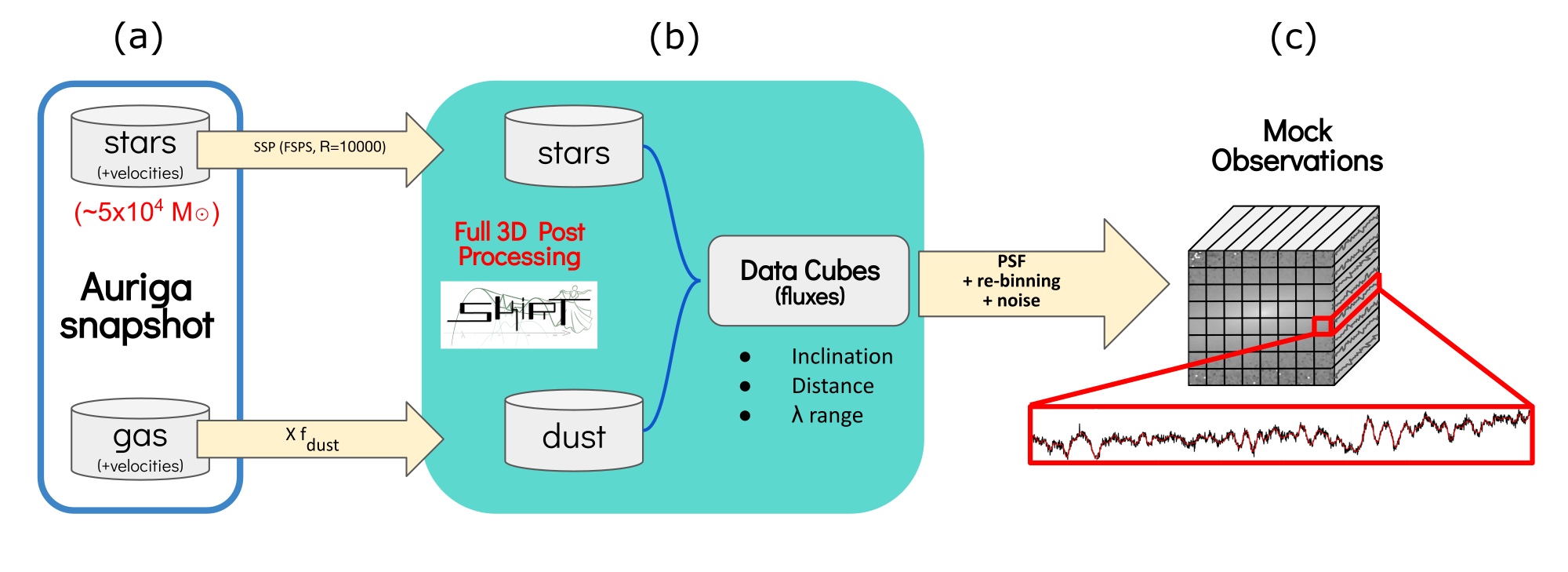}
  \caption{Scheme of the methodology used to create the synthetic observations. (a) Stellar particles and gas cell positions and velocities are taken from a particular snapshot in the Auriga output along with information relevant for the models used (i.e. age, metallicity, density, etc.). (b) Using \texttt{SKIRT}, a Monte Carlo simulation is performed where each particle is assigned a SSP from FSPS (C.Conroy, internal communication). The result from this step is an IFS-like data cube with spectra at each spatial position. The wavelength range and viewing angle of the synthetic observations are set in this step. (c) Further processing of the data cubes is performed by convolving spatially with a particular PSF, adding noise and re-binning the data to specific spatial and spectral resolution elements.}\label{fig:method_map}
\end{center}
\end{figure*}


\section{Post-processing scheme with SKIRT}\label{sec:post-pro}

\subsection{The \texttt{SKIRT} radiative transfer code}\label{sec:skirt}

To generate the synthetic spectroscopic observations for this study we use the public 3D radiative transfer code \texttt{SKIRT}~\citep[][]{Baes2011EfficientSkirt,Camps2015SKIRT:Architecture}\footnote{Available at \url{https://skirt.ugent.be}}. This code uses the Monte Carlo technique to simulate the physical processes that radiation experiences while propagating through dusty media, including scattering, absorption and thermal emission by dust. 

We use the most recent version of the code, \texttt{SKIRT 9}~\citep[][]{Camps2020SKIRTGrains}, in which kinematic effects both by the motion of the sources and the media can be included as an option in the simulations. This is implemented by storing the wavelength of each photon packet during the simulation and its changes due to Doppler shifts induced by the emitting sources and the interactions with the medium (re-emission and scattering by dust). This key feature provides us with a tool to derive the kinematic properties of the simulated galaxies taking into account the effects of the dust. 

In essence, \texttt{SKIRT} has three components, the sources of radiation, e.g. stars, star forming regions, AGN, etc., the media they propagate through, in our case the interstellar dust, and the instruments that record the radiation reaching a specific viewpoint. The code samples photon packets from the spatial and spectral distribution of the radiation sources and follows their paths between the source and the instruments, simulating the interactions with the media through a Monte Carlo simulation.

The code features multiple built-in 3D geometries for radiation sources and media~\citep{Baes2015SKIRT:Simulations} in addition to mechanisms for importing models generated by hydrodynamical simulations from particle- or grid-based representations. For particle-based imported sources, as are the stars in the Auriga simulations, the spectral distribution is selected from SED templates. \texttt{SKIRT} provides a library of templates ready to be used and also the choice of importing custom libraries. 
 
Different detector types permit the recording of fluxes in the spatial and spectral dimensions assuming a specific orientation of the source in the sky and a distance. In this work we use \texttt{SKIRT} to generate integral field spectroscopy (IFS) data cubes with the Auriga galaxies as our input models. We follow a procedure similar to the one described in~\citet{Kapoor2021High-resolutionProject} where, as illustrated in Fig.\ref{fig:method_map}, particle data are extracted from a simulation snapshot as input for a \texttt{SKIRT} run, which results in a simulated datacube that contains recorded fluxes. The main difference lies in the type of detector and wavelength range chosen for the output. In their work, the radiative transfer simulations generated photometry and images for various filters while in this work we store spectra at each pixel in our field of view (FOV). In the following sections we present a general description of the data extraction and simulation configuration, along with the main changes in the simulation configuration with respect to~\citet{Kapoor2021High-resolutionProject}. For a more in depth discussion about the choice of the model parameters see~\citet{Kapoor2021High-resolutionProject}.

\subsection{Input model extraction from Auriga snapshots}\label{sec:extract}

For our input models we use two components, an emission source given by the stellar distribution from Auriga and a medium, for the radiation to propagate through. Since the Auriga simulations do not include dust we adopt the interstellar gas as a tracer of the dust distribution. For each galaxy in our sample we extract the stellar particles and gas cells at a particular snapshot using a cubical aperture centered at the galaxy's center of mass and oriented using the momentum vector aligned with the z-axis. The length of the cube is set by a stellar surface mass density threshold of $0.2~{\text{M}}_\odot~{\text{pc}}^{-2}$; stellar particles and gas cells outside of this volume are ignored. Along with the 3D positions and velocities, additional information on the stellar and gas properties of each particle/cell is extracted to aid in the selection of models as explained in the following sections.

\subsubsection{Emission sources}\label{sec:em-sources}

Our Monte Carlo simulations follow the paths of randomly sampled photon packets emitted by the stellar particles from a particular Auriga galaxy. The wavelength of each of these photons is sampled from the spectral energy distribution, which in our case is a different single stellar population (SSP) spectrum depending on the mass, age and metallicity of the stellar particle. We use the FSPS-based models~\citep{Conroy2009TheGalaxies} but with the C3K high-resolution synthetic stellar spectra computed by C. Conroy (priv. comm), assuming a~\citet{Salpeter1955THE1954} IMF. The Auriga simulations use a Chabrier IMF, but this difference in choice of the IMF does not affect the spectra at a crucial level for the specific goal of this paper, investigating the effect of dust attenuation and instrumental characteristics on the observed stellar kinematics. We note that using a top-heavy IMF would change the spectra, with possible effects on the extracted kinematic and stellar population properties. The C3K templates offer a spectral resolution of $\sigma_{\rm{mod}}\sim$ 12 km s$^{-1}$ allowing for the computation of velocities at a high resolution. The age grid for these templates goes from $\log_{10}$(age yr$^{-1}$) $ = 5.0$ to $\log_{10}$(age yr$^{-1}$) $=10.3$ with a spacing of $\Delta\log_{10}$(age yr$^{-1}$) $ = 0.05$ and the metallicities available are Z = 0.0001, 0.0003, 0.0004, 0.0008, 0.0014, 0.0025, 0.0045, 0.008, 0.0142, 0.0253, 0.0449. The code interpolates between these templates to provide a model for the specific age and metallicty of each stellar particle. 
In this work we are focusing on the stellar kinematics of the simulated galaxies and therefore we do not add and extra emission source for those stellar particles young enough to be considered star forming regions as other works with similar procedures have done~\citep{Kapoor2021High-resolutionProject, Camps2016Far-infraredSimulations, Trcka2022UVGalaxies}.

\subsubsection{Diffuse dust distribution}\label{sec:medium}

We trace the dust distribution from the gas in the simulation, assuming that a constant fraction of the metals in the ISM is in the form of dust grains. In order to separate the cool interstellar gas from the hot circumgalactic medium, we only assign dust mass to gas cells that satisfy the temperature-density condition proposed by~\citet{Torrey2012Moving-meshDiscs}:
\begin{equation}\label{ec:dust-alloc}
    \log_{10}\left( \frac{T}{K}\right) < 6 + 0.25 \log_{10} \left( \frac{\rho}{10^{10} h^2\rm{M}_{\odot} \rm{kpc}^{-3}}\right).
\end{equation}

Different partitioning schemes have been used in previous studies to allocate the dust-containing ISM, and we refer the reader to~\citet{Kapoor2021High-resolutionProject} and~\citet{Trcka2022UVGalaxies} for a more thorough discussion. Figures 2 and 3 of these papers, respectively, show the impact of these choices on the resulting dust distribution. We use the current prescription as it belongs to one of~\citet{Kapoor2021High-resolutionProject} fiducial models. Dust is allocated to the cells that meet the condition in (\ref{ec:dust-alloc}) as $\rho_{\rm dust} = f_{\rm dust} Z \rho$, where $\rho$ and Z, the local density and metallicity of the gas cell, are taken from the Auriga snapshot and $f_{\rm dust}$, the dust-to-metal ratio is chosen to be $f_{\rm dust} = 0.14$. The choice in value for $f_{\rm dust}$ was taken from~\cite{Kapoor2021High-resolutionProject}, where this free parameter of the model was calibrated through comparisons with the observational DustPedia galaxy sample~\citep{Davies2017DustPedia:Universe} using colour-colour and other broad-band luminosity correlations.   

The dust density distribution is discretized using an octree structure computed by \texttt{SKIRT}~\citep{Saftly2013UsingSimulations,Saftly2014HierarchicalSimulations} from the original Voronoi mesh. The parameters used in the construction of the octree are the same as used in~\citet{Kapoor2021High-resolutionProject}. For the dust model our simulations use the \texttt{THEMIS}~\citep{Jones2017TheTHEMIS} dust mixture that contain silicates and hydrocarbon particles and we select 15 grain size bins per population. 

\subsection{Radiative transfer simulation parameters}\label{sec:sim-params}
Radiative transfer simulations are performed in extinction-only mode since we are only interested in the optical wavelength range; the re-emission of absorbed flux by dust in the infrared is not needed. This results in a reduction of the simulations computational time.

Because \texttt{SKIRT} uses a Monte Carlo technique to perform the radiative transfer in the simulations the resulting fluxes will have a noise associated to them related to the number of photon packets used.  We run our simulations using $5\times10^9$ photon packets. This choice was made by considering two statistical diagnostic quantities, the relative error, R, and the variance of the variance, VOV, as defined in~\cite{Camps2020SKIRTGrains}. 

As described in~\cite{Camps2020SKIRTGrains}, for a simulated flux to be reliable it must have values of $\rm{R} \leq 0.1$ and a $\rm{VOV} \leq 0.1$. This condition is not matched everywhere in the data cubes we generate, so spaxels that do not meet this condition were not considered in the analysis. However, all spaxels within 1 R$_e$ of the centre of the galaxies are within the confidence interval and in most cases it covers the area inside $\sim 2 \rm{R}_e$.

The goal of this work is to produce synthetic spatially resolved spectroscopic observations of the Auriga galaxies. To do this, we make use of the \texttt{FullInstrument} object within SKIRT, which allows for the recording of the surface brightness at every pixel, of a given field of view, and each wavelength, within a given range. The result is an IFS-like data cube that stores the spectra for each spaxel. The specific wavelength range, spatial extent and resolutions are dependent on the target observational sample to mock and therefore these parameter selections will be described in the next section.

We use the \texttt{SKIRT} functionality that allows for the separate recording of the flux components in our simulations. Apart from the standard data cube that fully accounts for dust attenuation, a “transparent” data cube is generated by this option, where the light coming from the stellar particles arrives at the instrument unaltered (it does not interact with the dust). This unattenuated data-set we use as "ground truth" in the comparisons ahead.

\begin{figure*}
\begin{center}
     \includegraphics[width = 
     \textwidth]{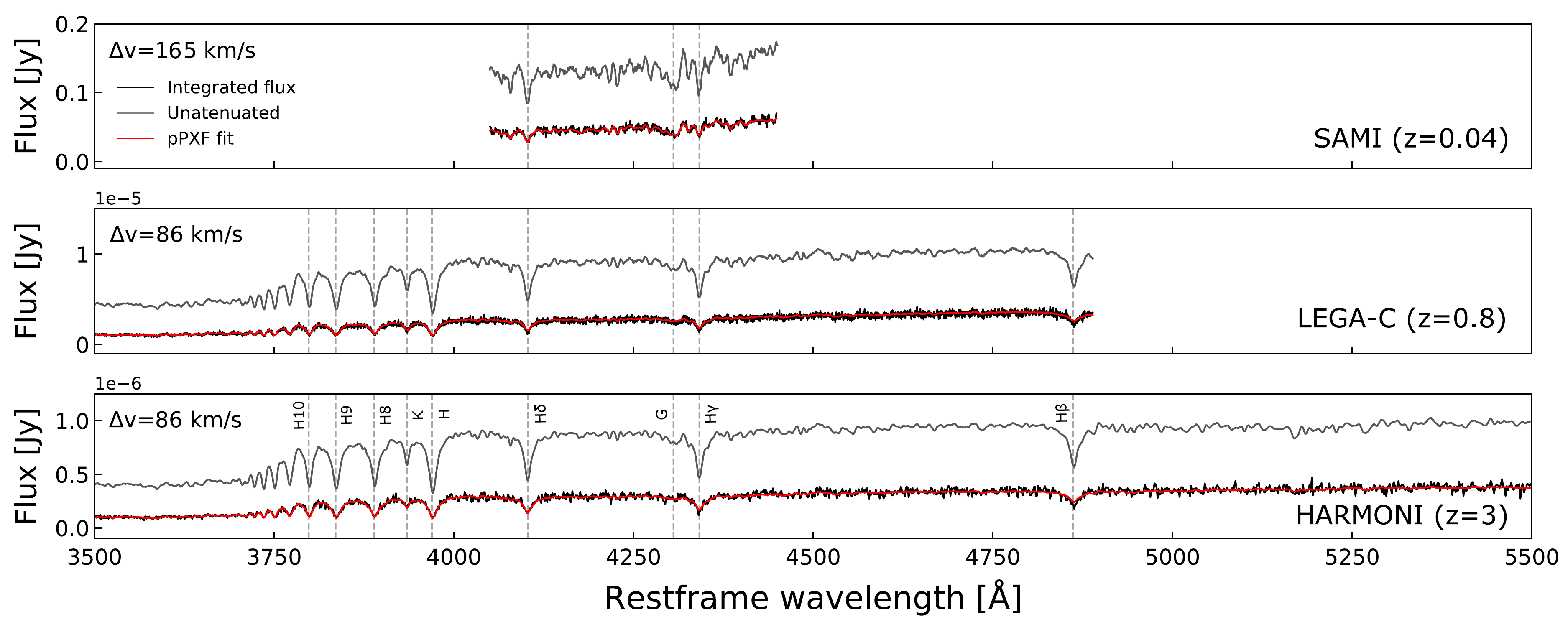}
     
  \caption{Integrated synthetic spectra for the SAMI-like (top), LEGA-C-like (middle) and HARMONI-like (bottom) simulations of haloes Au-6 at z=0, Au-6 at z=0.8 and Au-29 at z=3 from the Auriga sample. The intrinsic unattenuated stellar spectra is shown in grey. In black, is the spectra after post-processing, this includes noise addition and seeing convolution. Red lines indicate the best-fit model obtained with \texttt{pPXF}. Absorption features relevant for the kinematics analysis are highlighted. }\label{fig:spec-range}
\end{center}
\end{figure*}


\begin{figure*}
\begin{center}
     \includegraphics[width = \textwidth]{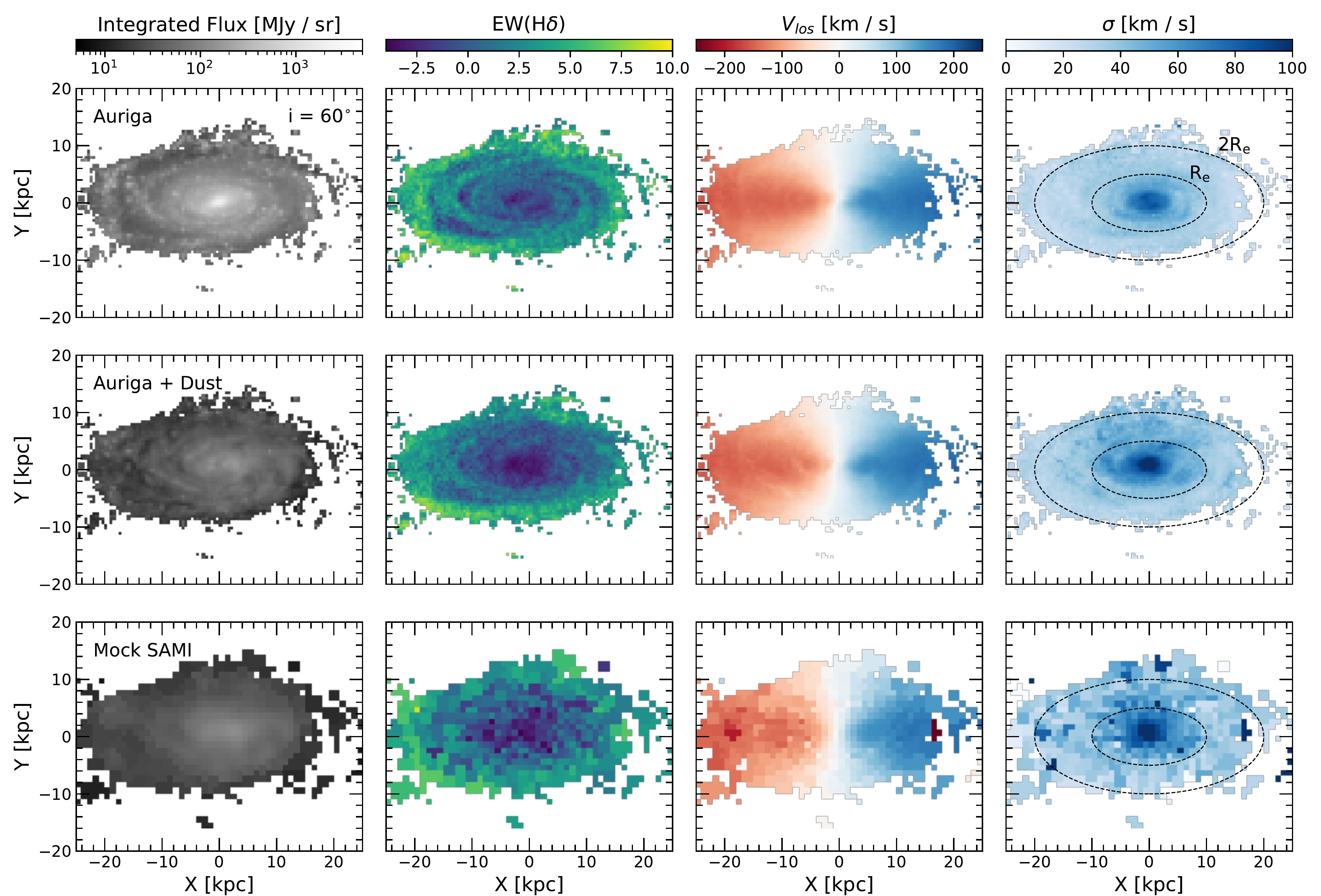}
  \caption{Example of our SAMI-like synthetic observations for halo Au-6 in the Auriga sample. The top row shows the unattenuated flux (a), $\rm{H\delta}$ equivalent width (b), line-of-sight velocity (c) and velocity dispersion (d) distributions at an inclination of i $=60\degree$. In the middle row are the maps for the post-processed version of the data cubes and the bottom row shows the results after the \texttt{SKIRT} run and further processing i.e. PSF convolution and noise addition. Ellipses on the last column (d) denote 1 $R_{\text{e}}$ and 2 $R_{\text{e}}$, derived using the particle data in~\citet{Grand2017TheTime}. Only 'reliable' pixels, as defined by the R statistic described in Section~\ref{sec:sim-params} are shown and used in the analysis.}\label{fig:sami_summary}
\end{center}
\end{figure*}

\section{Synthetic stellar kinematics and comparison to the SAMI Galaxy Survey}\label{sec:final-steps}

In this section we describe the methods used to convert the resulting \texttt{SKIRT} data cubes into the synthetic observations mimicking the specifications of the SAMI Galaxy Survey of present-day galaxies. We also present the kinematic analysis performed on the synthetic observations along with comparisons between the global kinematics obtained from the simulated data using our mock observations and those measured from the SAMI Galaxy Survey spectra.  

\subsection{The SAMI Galaxy Survey}\label{sec:sami}

The SAMI Galaxy Survey~\citet{Croom2012TheSpectrograph}, performed using the 3.9 m Anglo-Australian Telescope, observed 2964 galaxies between $0.004 < z < 0.095$, this includes 2083 targets from the Galaxy and Mass Assembly (GAMA) G09, G12 and G15 regions~\citep{Driver2011GalaxyRelease} and 881 from 8 clusters presented in~\citet{Owers2017TheProperties}, resulting in a sample with a broad range of environments and galaxy stellar mass ($\rm{M}_* \sim 10^8$ -- $10^{12} \rm{M}_{\odot}$). SAMI is a multi-object integral field spectrograph consisting of 13 hexabundles, sets of 61 individual optical fibres which feed into the AAOmega spectrograph~\citep{Saunders2004AAOmega:Overview,Smith2004AAOmega:AAT,Sharp2006PerformanceSpectrograph}. Each of these fibres has a diameter of $1\farcs6$ resulting in a $\sim 15\arcsec$ diameter region on the sky per bundle. 

Each galaxy in the survey is observed using both the blue (3750--5750\AA) and red (6300--7400\AA) arms of the spectrograph at spectral resolutions of $R=1808$ and $R=4304$ respectively~\citep{Croom2021TheRelease}. 

As reported in~\citet{Allen2015TheRelease}, the point spread function (PSF) for the survey varies from observation to observation and is primarily determined by the atmospheric conditions at the telescope. The distribution of PSF full width at half maximums (FWHM) measured on the final cubes of the survey vary within the range 1.4--3.0 arcsec with a median value of 2.1 arcsec.

The redshift of the observations in the SAMI Galaxy survey ranges from 0.004 < z < 0.095, with a median value of z$\sim$0.04, resulting in physical sizes of $\sim 12$ kpc for the FOV diameter and $\sim 1.3$ kpc for each optical fibre at this distance.

\subsection{Original SKIRT setup}\label{sec:sami-skirt-setup}

To recreate the specifications mentioned above in our mocks we choose a field of view of $50\times 50$ kpc$^2$ where the major axis of the galaxy is aligned with the x-coordinate in the sky. In this configuration each pixel is a square with a side of 0.5 kpc width. This serves as the base spatial sampling. We note that the SAMI FOV typically covers a smaller physical area than chosen for our simulations, however at this redshift it should reach at least 1 R$_e$, the area where most of our analysis is based on. The \texttt{SKIRT} simulations are run in "local universe mode", where the instrument stores the resulting fluxes on the rest-frame wavelength. In this case the wavelength range chosen spans $4050 < \lambda < 4450$ \AA~with a spacing of $\Delta v =$ 15 km s$^{-1}$. This range was chosen to be shorter than the SAMI observations to reduce the simulation time needed. However the range chosen contains important absorption features necessary for the extraction of the stellar kinematics, including $\rm{H}\delta$, $\rm{H}\gamma$, the G band at 4303.61 and various Fe features, as can be seen in the top panel of Figure~\ref{fig:spec-range}, which shows the integrated spectrum for one of the galaxies in our sample.

To examine the inclination dependence on the observed kinematics, we generate data cubes for a series of inclination angles: $i=25.84\degree, 45.57\degree, 60\degree, 72.54\degree$. These were chosen to match some of the inclinations included in the ~\citet{Kapoor2021High-resolutionProject} final broadband imaging catalogue. The azimuth was arbitrarily chosen at $\phi=142.47\degree$ to match one of the orientations in the mock photometry in~\citet{Kapoor2021High-resolutionProject}. We also added instruments with azimuth values $\phi=0\degree$ and $\phi=90\degree$ at the $i=72\degree$ view, to assess the dependence of our results on the specific geometry of the simulated galaxies.

We generate these mock data cubes for a subsample of the original 30 Auriga galaxies, consisting of the ten halos 1, 3, 6, 7, 13, 14, 17, 28, 29 and 30. We chose this set to represent the variety in the Auriga sample (Fig.~\ref{fig:subsample}). As stated above, all galaxies have similar masses and SFR, as the original purpose of the Auriga project was to produce Milky Way analogs. However, our sample contains several bulge-dominated systems, such as haloes 13, 17 and 30.

\subsection{Addition of observational effects}\label{sec:obs-effects}

The first step into the process of adding instrumental effects to our synthetic data cubes is to match the spectral resolution of the SAMI Galaxy Survey data. We start by convolving the spectra at each spaxel with the line-spread-function (LSF). We use a Gaussian kernel with FWHM set by the square root of the quadratic difference of the C3K and SAMI spectra FWHM. For SAMI we assume FWHM = 165 km/s, the resolution of SAMI's blue arm as reported in~\citet{Croom2021TheRelease}.

Then we proceed by degrading the spatial resolution of each resulting~\texttt{SKIRT} data-cube. We emulate the seeing effect by convolving the images at each wavelength slice of our data cubes with a Moffat function~\citep{Moffat1969APhotometry}. The profile is determined by two parameters, $\alpha$, the core width and $\beta$, the power that depend on the seeing following $\rm{FWHM} = 2\alpha\sqrt{2^{1/\beta} - 1}$. We use a fixed $\beta=4.765$~\citep{Trujillo2001ThePSF} and a FWHM=2 arcsec, which at the chosen redshift of z=0.04 represents a physical size of 1.64 kpc. We also re-bin each spatial element to twice its side, resulting in physical sizes of 1 kpc per spaxel.

Lastly, we emulate the signal-to-noise conditions of the survey by perturbing the fluxes in our cubes following a Gaussian noise. We use a constant noise value across the cube. In combination with the simulated fluxes this results in a distribution of signal to noise where the central brightest pixels have higher S/N values than the dimmer ones on the outskirts. We determine the noise level assuming a fixed signal-to-noise at 1R$_e$ of S/N $= 5$\AA$^{-1}$.

On actual spectroscopic observations, we expect to see a wavelength dependence on the noise level due to the varying sensitivity of the detector and spectrographs. Even within each arm of the SAMI instrument the detector response fluctuates at the edges which translates to a higher noise~\citep{Bryant2015TheSelection, vandeSande2017TheKinematics}. We note that the spectral range of our mock spectra is much narrower than that used in the survey and corresponds only to a portion of the coverage of the blue arm, so we chose to use a constant level of noise across wavelength for simplicity purposes. We expect that including a more realistic noise, derived from the observations or from detector response models~\citep{Ibarra-Medel2019OpticalMethod}, would contribute to the uncertainty in the measured kinematics. The addition of different noise models is easy to implement, and it is the approach we follow for the generation of our LEGA-C long-slit mocks (see Section~\ref{sec:mock-legac}).

\subsection{Spectral fitting}\label{sec:spec-fit-sami}

Once we have the final version of our synthetic cubes we measure line-of-sight kinematics and absorption line indexes. In the outskirts of our selected FOV, S/N is too low for the spectra to be used individually. To alleviate this issue, analogous to observed IFS data, we bin spaxels to reach a minimum of S/N$=3$\AA$^{-1}$. We perform this using the~\citet{Cappellari2003AdaptiveTessellations} Voronoi tesselation method implemented in the \texttt{VorBin}\footnote{Available at \url{https://www-astro.physics.ox.ac.uk/~cappellari/software/}} package.

To recover the stellar kinematics we used the penalized pixel fitting code \texttt{pPXF}~\citep{Cappellari2004ParametricLikelihood,Cappellari2017ImprovingFunctions}\footnote{Available at \url{https://www-astro.physics.ox.ac.uk/~cappellari/software/}} to fit a Gaussian line of sight velocity distribution (LOSVD) at each spaxel in the data cube. We use the same high-resolution synthetic SSP templates from FSPS that were used to generate the spectra to perform the fitting. We fit the first two moments of the LOSVD, the line of sight velocity, $v_{\rm{los}}$, and velocity dispersion, $\sigma$, and use a polynomial of degree 12 to fit the continuum. We follow a procedure similar to that described in~\citet{vandeSande2017TheKinematics}, where at a first stage we create annular binned spectra of higher S/N to derive local optimal templates. We select 5 elliptical annular bins following the orientation of the light distribution and for each bin derive the mean flux of the spaxels within it. For each of the 5 spectra we perform a fit, which will result in 5 sets of optimal templates at each annular bin. Then, each spaxel is fitted using the set of templates corresponding to the annulus the spaxel belongs to. As a result we obtain $v_{\rm{los}}$ and $\sigma$ maps at all orientations and for all the components described previously. Columns 3 and 4 of Figure~\ref{fig:sami_summary} show the LOSVD maps for both the original \texttt{SKIRT} output datacube and the resulting SAMI-like data. An intermediate step, with the \texttt{SKIRT} post-processed cube but without instrumental effects applied is also shown. There are subtle differences in the velocity distributions for the unattenuated and attenuated cases but in general they seem to agree. The largest loss of structure happens for the fully mocked data, due to the noise addition and degradation in spatial resolution. The most noticeable changes in the 2D kinematics maps are in the velocity dispersion distributions. As a result of the post-processing the dispersion values at the centre of the galaxy become higher. The maximum velocity dispersion in the central pixels reaches a value ~15 km s$^{-1}$ higher for the fully mocked observations compared to the original model. In this case the spatial binning of the spaxels contributes to the offset in velocity dispersion. We also measure the equivalent width of the H$\delta$ Balmer absorption, EW($\rm{H\delta_A}$), at each spaxel on the cube. We use the definition in~\citet{Worthey1997HPopulations}, and use the wavelength range 4083.50 -- 4122.25 \AA~for the line index. The resulting maps are shown in the second column of Figure~\ref{fig:sami_summary} where all maps show a radial trend, with higher values at the outskirts compared to the centre. The addition of dust results in some loss of the spiral arm structure traced by the unattenuated EW measurements, specially at the centre. This structure is not visible on the lower panel, where the spatial resolution is degraded.

\subsection{Global kinematics}\label{sec:global-kin}

To quantify the effects that spatial resolution, instrumental effects, and dust attenuation have on the synthetic spectra we compute the global stellar kinematics within 1 $R_{\text{e}}$. The radius chosen corresponds to $R_{\text{e}}=1.68\ r_{\text{d}}$, where $r_{\text{d}}$, the exponential radius is taken from~\citet{Grand2017TheTime}, where a Sérsic and exponential profile are fitted simultaneously to the stellar surface density profile of each halo. We select the pixels within this radius in our~\texttt{pPXF} resulting maps and compute the rotation velocity, $V_{\rm{rot}}$, as
\begin{equation}\label{ec:vrot}
    V_{\rm{rot}} = \frac{W}{2\,(1+z)\sin(i)},
\end{equation}
 where $W$ corresponds to the velocity width, the difference between the 90th and 10th percentiles of the histogram of rotation velocities ($W=V_{90} - V_{10}$). The inclination, $i$, corresponds to the inclination chosen for each instrument in \texttt{SKIRT} and $z$ is a fixed value for all halos set on the simulation. We also measure the velocity dispersion, $\sigma_{\text{e}}$ as,
 
 \begin{equation}\label{ec:sigma}
     \sigma_{\text{e}}^2 = \frac{\sum_j F_j \sigma_j^2}{\sum_j F_j},
 \end{equation}

where $\sigma_i$ corresponds to the velocity dispersion in each spaxel and $F_i$ to the corresponding continuum flux. 

In addition to these measurements, we also measure line-of-sight kinematics for spectra produced using only partial elements of the mock process. These will help us better understand the effects of adding instrumentals and dust.  

These extra versions of the mock spectra correspond to the following:

\begin{itemize}
    \item "Intrinsic": Mock data-cubes generated with SKIRT (without radiative transfer) at a high spectral ($\Delta v=15$ km s$^{-1}$) and spatial resolution ($\Delta x= 0.5$ kpc). These mocks are oriented at a face-on ($i=0$\degree) and edge-on ($i=90$\degree) view of the galaxy disk. 

    \item "Dust only": Data-cubes generated with SKIRT (with radiative transfer) at the native resolution (none of the instrumentals are added). At all orientations in the original run. 
    
    \item "Instrumentals only": Data-cubes generated with SKIRT (without radiative transfer) with all instrumentals added, including LSF and PSF convolution, binning and noise addition. At all orientations in the original run. 
    
\end{itemize}

Global kinematics are also measured from these datacubes. From here on out we will refer to the rotation velocity, $V_{\rm{rot}}$, from the "intrinsic" edge-on mock and the velocity dispersion, $\sigma_{\text{e}}$, from the "intrinsic" face-on mock as the "ground truth" kinematics for each galaxy.  

\begin{figure}
\begin{center}
     \includegraphics[width =\columnwidth]{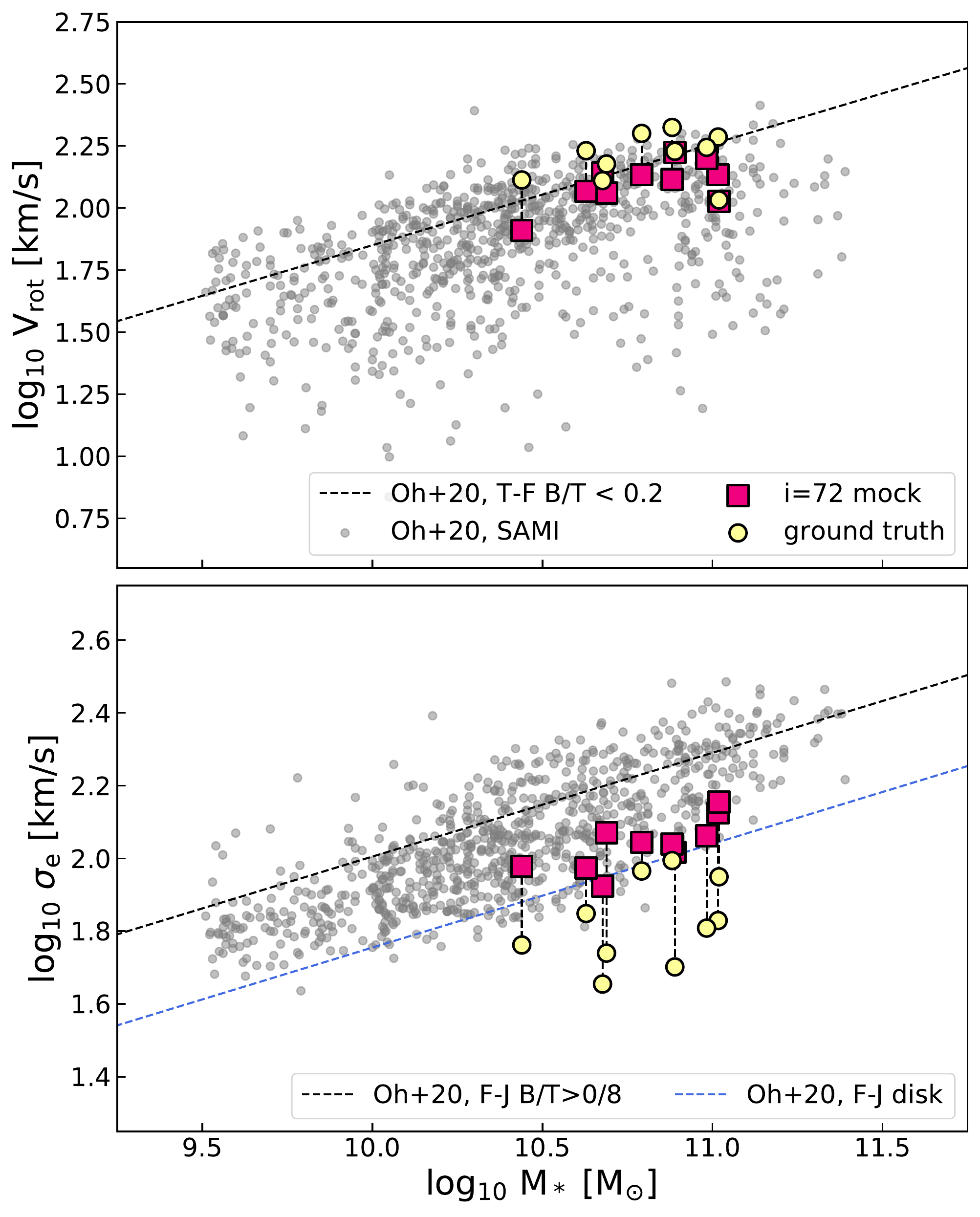}
  \caption{Comparison of observed vs simulated scaling relations with stellar mass. The stellar mass Tully-Fisher (top) and Faber-Jackson (bottom) relations as observed for the galaxies in the SAMI Galaxy Survey~\citep{Croom2021TheRelease}. Individual galaxies in the survey are shown in grey while magenta squares indicate the recovered kinematics from the synthetic observations of a sub-sample of the Auriga galaxies from this work. The inclination of the observations displayed correspond to $i=72$\degree, the highest inclination used. The ground truth values, corresponding to edge-on rotation velocity and face-on velocity dispersion from unattenuated, unperturbed data are shown as yellow circles. Black, dashed lines indicate the best-fit relations calculated by~\citet{Oh2020TheDisks} using galaxies with $B/T < 0.2$ and $B/T > 0.8$ for T-F and F-J respectively. There is good agreement between our Auriga derived kinematics and the results from the SAMI Galaxy survey, however offsets with the ground truth values can be observed, specially for velocity dispersion. All our velocity dispersion measurements fall below the F-J relation (calculated only using bulge dominated systems) and are closer to the best-fit relation recovered by~\citet{Oh2020TheDisks} using only the disk components resulting from their disk/bulge decomposition (in blue).}\label{fig:corr-compare}
\end{center}
\end{figure}

\subsection{Comparison with SAMI Galaxy Survey observational data}\label{sec:sami-compare}

We start our analysis by comparing the properties derived from our synthetic SAMI Galaxy Survey data with the actual observations obtained in the survey~\citep{Croom2021TheRelease}. In Figure~\ref{fig:corr-compare}, we compare the positions of the global kinematics derived from our $i=72$\degree~realisation in the $M_\star$--$V_{\text{rot}}$ (top panel) and $M_\star$--$\sigma_{\text{e}}$ (bottom panel) spaces with respect to the SAMI Galaxy Survey derived quantities from~\citet{Oh2020TheDisks}. As shown in the top panel, the rotation velocities derived from our synthetic data are consistent with those observed in disk galaxies in the SAMI Galaxy Survey, as shown by the agreement between our data points and the fiducial stellar mass Tully-Fisher relation derived by~\citet{Oh2020TheDisks} using single-component kinematic fits to late-type galaxies with D/T < 0.2. When we consider the values retrieved from our "ground truth" realisation, these are still consistent with the SAMI sample, however, we see a systematic offset towards lower values in the fully processed sample, produced by the addition of dust and instrumental systematics added to the simulation data.  

Velocity dispersions, on the other hand, while still consistent with the distribution of the SAMI sample fall below the derived Faber-Jackson relation by~\citet{Oh2020TheDisks}. Their fiducial relation was computed using bulge-dominated galaxies with B/T > 0.8, a condition that none of the galaxies in the Auriga sample at z=0 holds. The galaxy with the lowest D/T value, Au-11 with D/T =0.29 was excluded from our selection due to its ongoing interaction with another system which would complicate our analysis of its kinematics. We still see a correlation between our derived velocity dispersions and stellar mass with a similar slope to the fiducial derivation from~\citet{Oh2020TheDisks}, but offset downwards, closer to the Faber-Jackson relation they derived using only the disk component from their disk-bulge kinematic decomposition. So the tension between the best-fit relation and the data derived from our synthetic observations can be attributed to the fact that, as mentioned before, galaxies in the Auriga sample are designed to resemble the Milky Way, resulting in a sample of disc-dominated galaxies. 

However, when we compare to the ground truth data we see a stronger disagreement between the data sets. Intrinsic velocity dispersions are lower than the ones derived from fully post-processed data by around 0.2 dex, and falling outside the observed distribution of $\sigma_{\text{e}}$ in the SAMI Galaxy Survey. It had already been shown in~\citet{VanDeSande2019TheSimulations} how multiple cosmological simulations consistently under-predicted velocity dispersion at fixed stellar mass. In their work they measure offsets ranging from 0.14 to 0.23 dex between the median velocity dispersion from the SAMI Galaxy Surveys and the simulated data from the \texttt{EAGLE$^+$}, \texttt{HORIZON-AGN} and \texttt{MAGNETICUM} cosmological simulations. This means that the stellar kinematics produced by the Auriga models are in agreement with spatially resolved observations at $z=0$ from the SAMI Galaxy Survey, however to reach this agreement a correct dust treatment and implementation of the instrumental effects must be taken into account. Specifically, the measured $\sigma_{\text{e}}$ scales with galaxy stellar mass, following the Faber-Jackson relation, but only because it is dominated by beam-sheared rotation, not the intrinsic velocity dispersion.

We note that the stellar masses used for our synthetic sample come directly from the simulation (Table 1 in~\citet{Grand2017TheTime}) in contrast to the SAMI Galaxy Survey values, derived from $i$-band luminosities and $g-i$ colours~\citep{Taylor2011GalaxyEstimates}. This difference could introduce biases in the placement of the data with respect to the observed correlations. Nevertheless for the intrinsic Auriga $\sigma_{\text{e}}$ values to be consistent with the observed data a shift of $\sim$1 dex in stellar mass would be needed. In~\citet{Kapoor2021High-resolutionProject}, stellar mass estimates were obtained from synthetic photometry for the Auriga galaxies using SED fitting with \texttt{CIGALE}. In general, they found good agreement between their results and the particle-based stellar masses, so we conclude that the systematics in the kinematics due to the post-processing are playing a key role in the agreement of both samples.    

\begin{figure}
\begin{center}
     \includegraphics[width =\columnwidth]{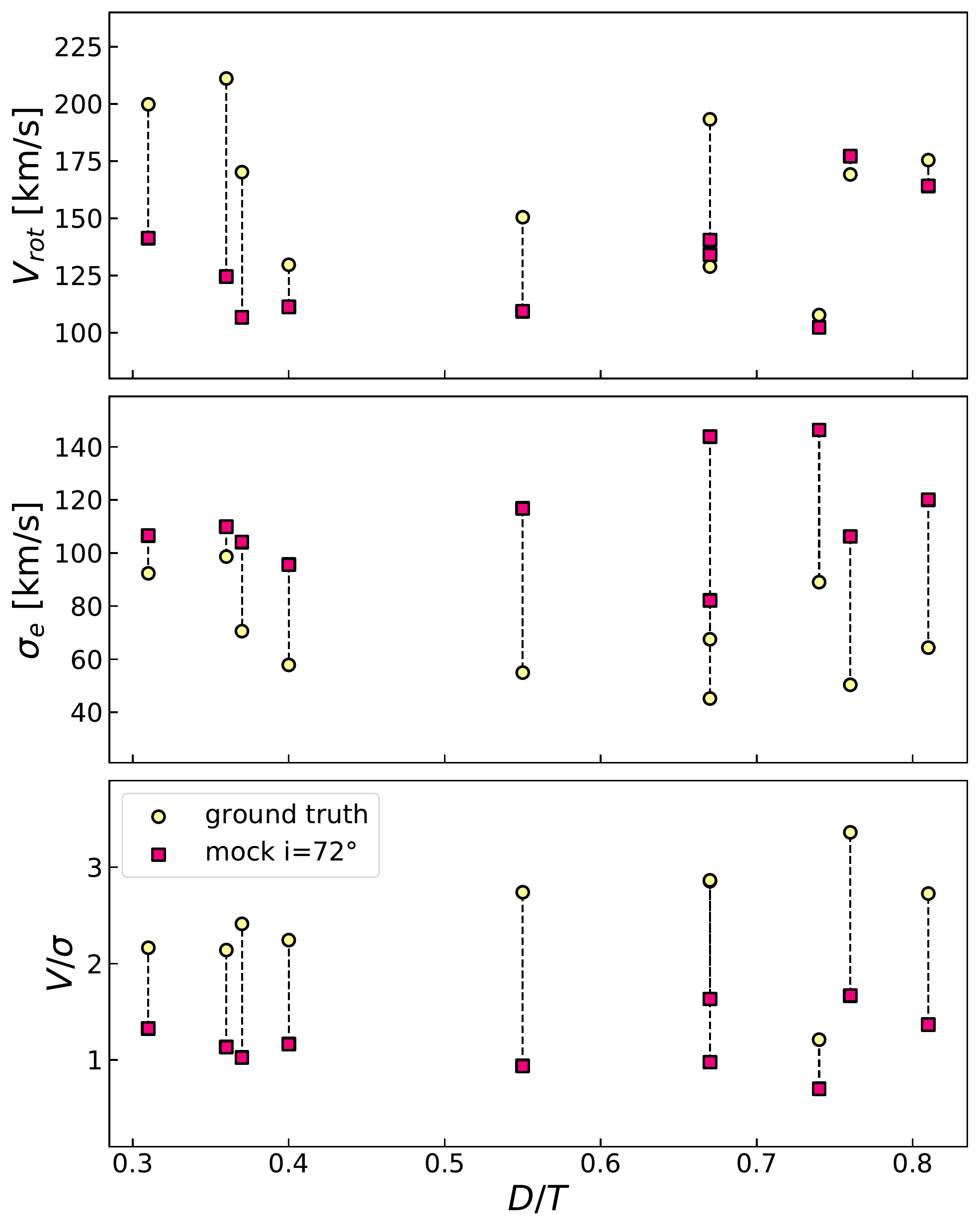}
  \caption{Global kinematic measurements and their dependency on the disk-to-total ratio. Rotation velocity (top), velocity dispersion (middle) and $V/\sigma$ (bottom) for our sub-sample of the Auriga simulated galaxies at $z=0$. Magenta squares indicate the values measured from the i=72\degree ~mocks of each galaxy in the sample. Yellow circles connected to the squares by a dashed vertical line correspond to the ground truth values calculated from unattenuated, unperturbed datacubes. In this case edge-on (i=90\degree) and face-on (i=0\degree) datacubes are used to measure rotation velocity and velocity dispersion, respectively. Values are displayed as a function of the disk-to-total ratio of the galaxy, calculated from a two-component fit to the light-profiles derived in~\citet{Grand2017TheTime}.}\label{fig:dt-trends}
\end{center}
\end{figure}

As for the scatter in the relations, both ground truth and mocked Auriga Tully-Fisher relations appear very tight, with little scatter in our sample, in contrast to the observed galaxies by~\citet{Oh2020TheDisks}. On the other hand, the relation between $\sigma_{\text{e}}$ and stellar mass has larger amounts of scatter, particularly for the ground truth case. The mock velocity dispersions measured at $i=72$\degree~show a tighter relation with reduced scatter. In~\citet{Oh2020TheDisks} it is shown that one source of scatter in both relations is the variation in morphologies of the galaxies, particularly the B/T is shown to correlate with the residuals of both Tully-Fisher and Faber-Jackson relations. We explore this possibility in Figure~\ref{fig:dt-trends}. We observe only faint correlations between the ground truth $\sigma_{\text{e}}$ and $V/\sigma$ with the disk-to-total ratio (D/T) measured in~\cite{Grand2017TheTime}. However, these correlations get washed away once the dust treatment and instrumentals are added, hinting at a differential effect of dust and PSF on the kinematics dependent on the morphology of the galaxies. It is worth noting that our sample is reduced in size and the range of values of D/T is much smaller than the one observed in the galaxies from the SAMI Galaxy Survey, thus limiting our analysis.

\begin{figure*}
\begin{center}
     \includegraphics[width =\textwidth]{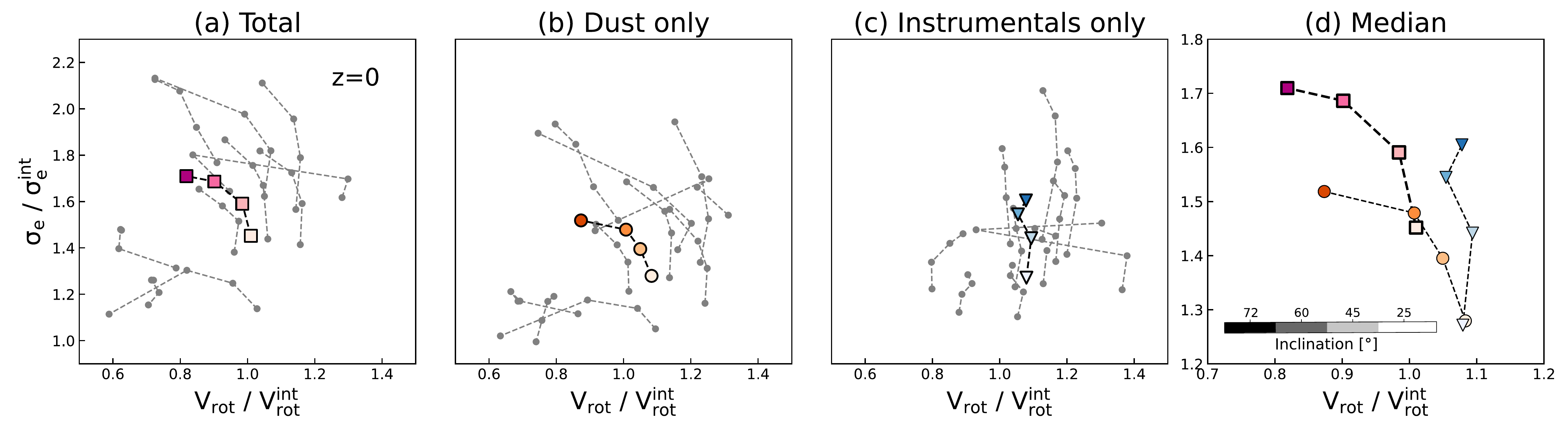}
  \caption{Effects of the post-processing on the recovered global kinematic properties for our sub-sample of the Auriga galaxies at $z=0$. Global kinematic properties were computed from our SAMI-like IFS data cubes and correspond to: rotation velocity (x-axis) and velocity dispersion (y-axis) within 1 $R_{\text{e}}$ of the centre of the galaxies. Properties are plotted as fractions of the "intrinsic" values, calculated using unattenuated data cubes with i=90\degree (edge-on) for the rotation and i=0\degree (face-on) for the velocity dispersion.
  For each individual galaxy in our selected sample, we generate synthetic spectra from 4 different viewing angles, at increasing inclination angles. The results from these realisations are plotted as grey dots connected by dashed lines in each panel. Each column shows a different step of the post-processing, from left to right: (a) the complete post-processing treatment including both radiative transfer and instrumental effects, (b) only post-processing with \texttt{SKIRT} at a high spatial resolution and (c) only PSF-convolution and noise addition but no radiative-transfer or dust attenuation of any kind.
  The median values per inclination are shown as, magenta squares in (a), orange circles in (b) and blue triangles in (c), with the shades of each colour going from lightest (i=25\degree) to darkest (i=72\degree) to indicate the inclination. In panel (d) we compare these trends highlighting the separate effects that dust and instrumentals have on the recovered global kinematics from our analysis. The rotation velocities presented are inclination corrected following Equation~\ref{ec:vrot}. }\label{fig:vrot-sigma}
\end{center}
\end{figure*}

\subsection{Disentangling dust attenuation and observational effects}\label{sec:dust-vs-instrumental}

We can further analyse our retrieved global kinematics and study the root of the observed offsets by separating the contributions of dust attenuation and instrumental effects. This is done by performing our analysis on partially processed data cubes, one with radiative transfer but at the native resolution of our Monte Carlo simulations and a second one without radiative transfer but including spectral and spatial convolution, pixel binning and noise addition. 

The results are shown in Figure~\ref{fig:vrot-sigma}. The left-most panel shows our fiducial run, including both radiative transfer with \texttt{SKIRT} and mimicking the instrumental specifications of the SAMI Galaxy Survey. Each grey curve shows the offsets in rotation velocity, $V_{\rm{rot}}/V_{\rm{rot}}^{int}$, and velocity dispersion, $\sigma_e/\sigma_e^{int}$ within 1 $R_{\text{e}}$ for a single Auriga galaxy in our sample. Connected by dashed lines, data points in each curve correspond to the increasing inclinations. We find that the offsets in both the rotational velocity and velocity dispersion values for our fully treated data cubes show a clear inclination dependency. Rotational velocity underestimations get larger as inclination increases while the resulting velocity dispersion is increasingly overestimated with respect to the ground truth line-of-sight velocity dispersion. 

Repeating our analysis on a set of data cubes that were post-processed with \texttt{SKIRT} but with no further degradation of the spectral and spatial resolution, we see a similar behaviour as in our fiducial run. This agreement is highlighted by the median trends in the right-most panel of Figure~\ref{fig:vrot-sigma}, where both trends evolve in the same direction however with different amplitudes in the offset values.

The trends with inclination can be attributed to the higher amounts of dust in the line-of-sight due to the inclination of the disk. A behaviour similar to this for the rotational velocity had already been reported on by~\citet{Baes2003RadiativeGalaxies}, where they analysed the effects of inclination and optical depth on the mean projected velocity and velocity dispersion of modeled disks. They show that even at small optical depths the effects of dust attenuation are very noticeable at high inclinations and interpret these discrepancies as the result of only capturing the radiation coming from the optically thin part of the galaxy. This would result in shallower rotation curves, tending towards a solid body rotation, that would translate to lower $V_{\rm{rot}}$ at the highest inclinations, even after applying the usual inclination correction term, $\sin(i)$, for the difference in velocity components.

On the other hand, the instrumental effects appear to affect mostly the velocity dispersions derived, an effect that also correlates with the inclination of the disk. This is a result of the limited resolution which leads to spatial mixing of radiation from stars with different velocities. This effect is most noticeable in the centre where the velocity changes between adjacent pixels are largest as shown in Figure~\ref{fig:sami_summary}. The trend is then a result of the face-on views of the disk having smaller velocity gradients and thus resulting in lower velocity dispersions when the PSF convolution blends the light, an effect that has been reported on by multiple studies~\citep{DEugenio2013Fast0.183, vandeSande2017TheSurveys, Harborne2020RecoveringData, Nevin2021AccurateKinematics}. 

\begin{figure*}
\begin{center}
     \includegraphics[width =\textwidth]{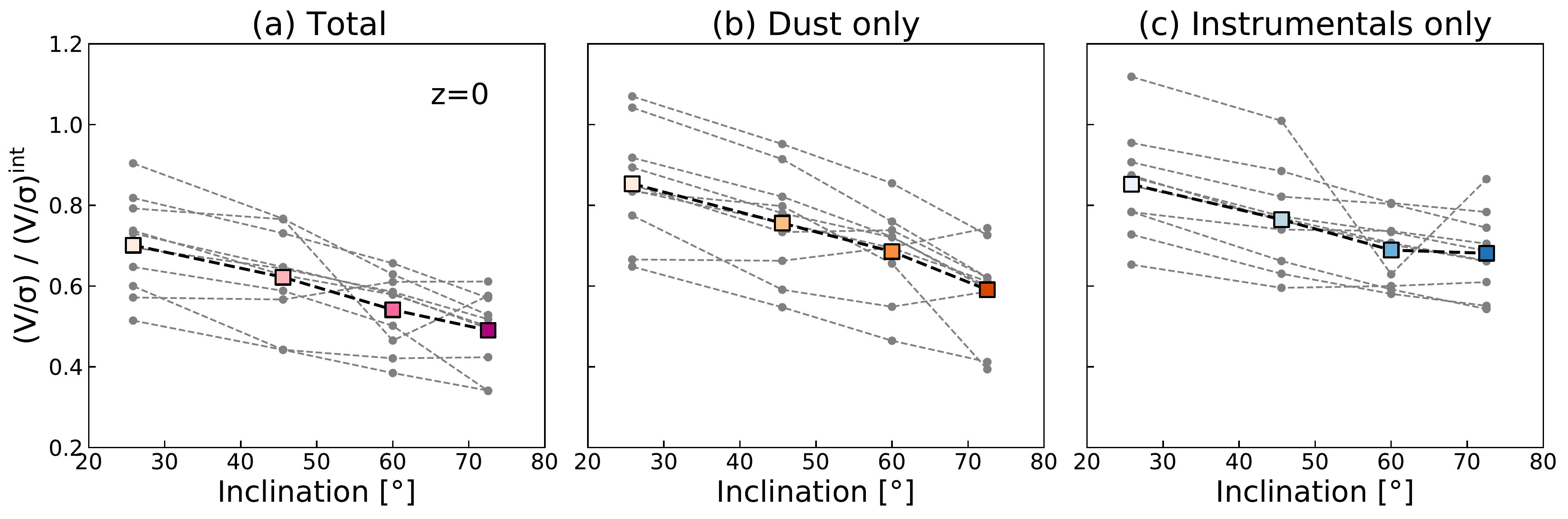}
  \caption{Effect of the post-processing on the inferred rotational support, $V/\sigma$, from the SAMI-like IFS data cubes as a function of the inclination of the synthetic observation. The y-axis shows the ratio between $V/\sigma$ within 1$R_{\text{e}}$, as recovered from our post-processed data cubes, and the "ground truth" we define as the rotational velocity and velocity dispersion calculated from unattenuated data cubes from an edge-on (i=90\degree) and face-on (i=0\degree) views of the galaxy. Values for individual galaxies in the sample are shown as grey dots joined by dashed lines, connecting the recovered values for the increasing inclination viewing angle used in the synthetic observations. Columns 2 and 3 separate the effects produced by dust attenuation (b) and instrumentals (c) to the recovered rotational support. Squares indicate the median value per inclination in the sample. On average the recovered rotational support is underestimated, an effect that becomes more prominent with inclination, reaching a factor $\sim 2$ at the highest inclination measured (i=72\degree).}\label{fig:rot-support-trends}
\end{center}
\end{figure*}

The combination of underestimating rotation velocities while overestimating velocity dispersion produces negative offsets in the recovered $V_{\rm{rot}}/\sigma_e$, as displayed on Figure~\ref{fig:rot-support-trends}. These offsets are more severe for the more inclined disks and the discrepancies can reach up to a factor $\sim2$ for our fully post-processed synthetic observations. This quantity is often used to measure the amount of ordered versus random motion in a system, so a systematic bias of this magnitude could lead to confusion in the classification of fast and slow rotators in a galaxy population.

At the resolution level of the SAMI Galaxy survey, both dust and observational systematics produce offsets of similar magnitudes. The discrepancies in rotational support are dominated by the overestimation of the velocity dispersion, which shows offsets of a higher magnitude than the line-of-sight velocity.

We can compare our results with those found in~\citet{Harborne2020RecoveringData}. In their study they generate mock kinematics for 25 modeled galaxies at different inclinations and seeing conditions and explore the dependence of $\lambda_{\text{R}}$ and $V/\sigma$ on the spatial resolution and other galaxy properties ($\epsilon$, n, R$_{\rm{eff}}$). For the range of sizes and resolution of our SAMI mocks, they find values between $\Delta (V/\sigma) = 0.9$ -- 0.7, depending on the projection and shape of the observed galaxies, which is consistent with what we recover for the run with only instrumental effects.

\section{High-Redshift Mock Continuum Spectroscopy}\label{sec:high-z}

In this section we further test our methodology to produce synthetic spatially resolved spectroscopy at higher $z$. We generate mock observations resembling the long-slit spectra from the LEGA-C survey at $z=0.8$ for a single galaxy and analyse the kinematics derived from them.  We also generate an IFS datacube following the instrumental characteristics of the future E-ELT's instrument HARMONI and simulate an observation for a source at $z=3$.

\subsection{Simulating LEGA-C Observations at z=0.8}\label{sec:z-0.8}
\subsubsection{The LEGA-C survey}\label{sec:lega-c}

The LEGA-C survey~\citep{vanderWel2016THEGyr, Straatman2018Field, vanderWel20210.6} is an ESO Public Spectroscopic survey with VLT/VIMOS. The survey targets $>$3000 $Ks$-band selected galaxies in the COSMOS field between $0.6 \leq z \leq 1.0$. The 20 hr long integrations result in high signal-to-noise stellar continuum detections of on average $\sim 19$ \AA$^{-1}$ at a spectral resolution of $R=3500$~in the range 6000 -- 9000~\AA. These conditions make the LEGA-C data an ideal sample to study the stellar kinematics of galaxies at high $z$. The field of view covered by the slits is 1.025 arcsec in width, in the wavelength direction,  and at least 8 arcsec long, with a pixel size of 0.205 arcsec (and 0.6~\AA~in wavelength).

\subsubsection{Generation of the mock spectroscopy}\label{sec:mock-legac}

In contrast to the sample at z=0, we only perform our high $z$ test using a single galaxy from the Auriga sample, Au-6. This galaxy was chosen arbitrarily with the only criteria being a sufficiently large stellar mass at the extraction snapshot corresponding to $z=0.8$. At this stage in the simulation Au-6 has a stellar mass of $\sim 1.6 \times 10^{10}$~M$_{\odot}$ and a dust mass fraction of $f_{\rm{dust}}=0.016$ (as estimated from our dust assignment procedure on the star forming cells in the simulation).  

To simulate the effects of the slit width on the resulting 2D spectra we first generate and IFS-like data cube that at a later step is stacked in the wavelength direction to obtain a mock 2D slit observation. At $z\sim 0.8$ the FOV used is $30.4\times8$ kpc and each pixel has a size of 1.6 kpc, this results in a data cube with 19 spaxels of length and a width of 5 spaxels. 

The wavelength grid used for these simulations spans the $6300 < \lambda < 9000$ \AA~range, at the observed frame, with a resolution of $\Delta v=$ 86 km s$^{-1}$. At $z=0.8$, the rest-frame wavelength range covers a wealth of absorption features useful for the kinematic analysis. Among them are lines in the Balmer series from $\rm{H}\beta$ at 4861.35 \AA~to H10 at 3798.0 \AA, and also include the CaK (3934.7 \AA), CaH (3969.5 \AA) and G band (4304.6 \AA) absorption features useful for age and metallicity determinations. An example of the resulting integrated spectra is shown in the middle panel of Fig.~\ref{fig:spec-range}, where both the unattenuated (grey) and attenuated (black) spectra are shown.

We choose a set of inclinations $i=25\degree, 60\degree, 80\degree,$ and $
90\degree$ where the major axis of the galaxy is aligned with the slit. We also add an observation with inclination $i=60\degree$ and $\omega=20\degree$, where the variation in roll angle is meant to explore the effects of miss-alignment between the slit and galaxy major axis. 

Before going from the 3-dimensional data cubes to 2-dimensional spectra we perform a PSF convolution to match the typical resolution of the survey at $z\sim 0.8$. We adopt a 2D Moffat profile with a FWHM=7.5 kpc ($\sim$ 1 arcsec) and a fixed $\beta=765$. Then we combine the 5 pixel rows representing the width of the slit by shifting the wavelength of the spaxel depending on the corresponding distance to the centre of the slit to simulate the effect of the light as it projects on the CCD. 

Once we have a 2D spectrum that matches the dimensions of the LEGA-C data, the final step is the addition of realistic noise. In this case the assumption of constant noise across wavelength range is not valid. Compared to our SAMI mocks the wavelength range selected for LEGA-C is much larger and also at the observed range the amount of telluric absorption and emission is higher at these longer wavelengths, making the noise highly dependent on wavelength. We select a real LEGA-C spectrum (ID:123292) matched in redshift and approximate stellar mass, and extract its 1D noise curve. We then compute an extra normalisation of the noise spectrum by setting the S/N of the central 5 pixels combined to a S/N$=20$\AA$^{-1}$. The perturbed spectrum is then created by drawing from a Gaussian distribution centred at the flux in our mock and a standard deviation given by the noise level at the corresponding sampled wavelength in the noise spectrum.  

\begin{figure*}
\begin{center}
     \includegraphics[width =\textwidth]{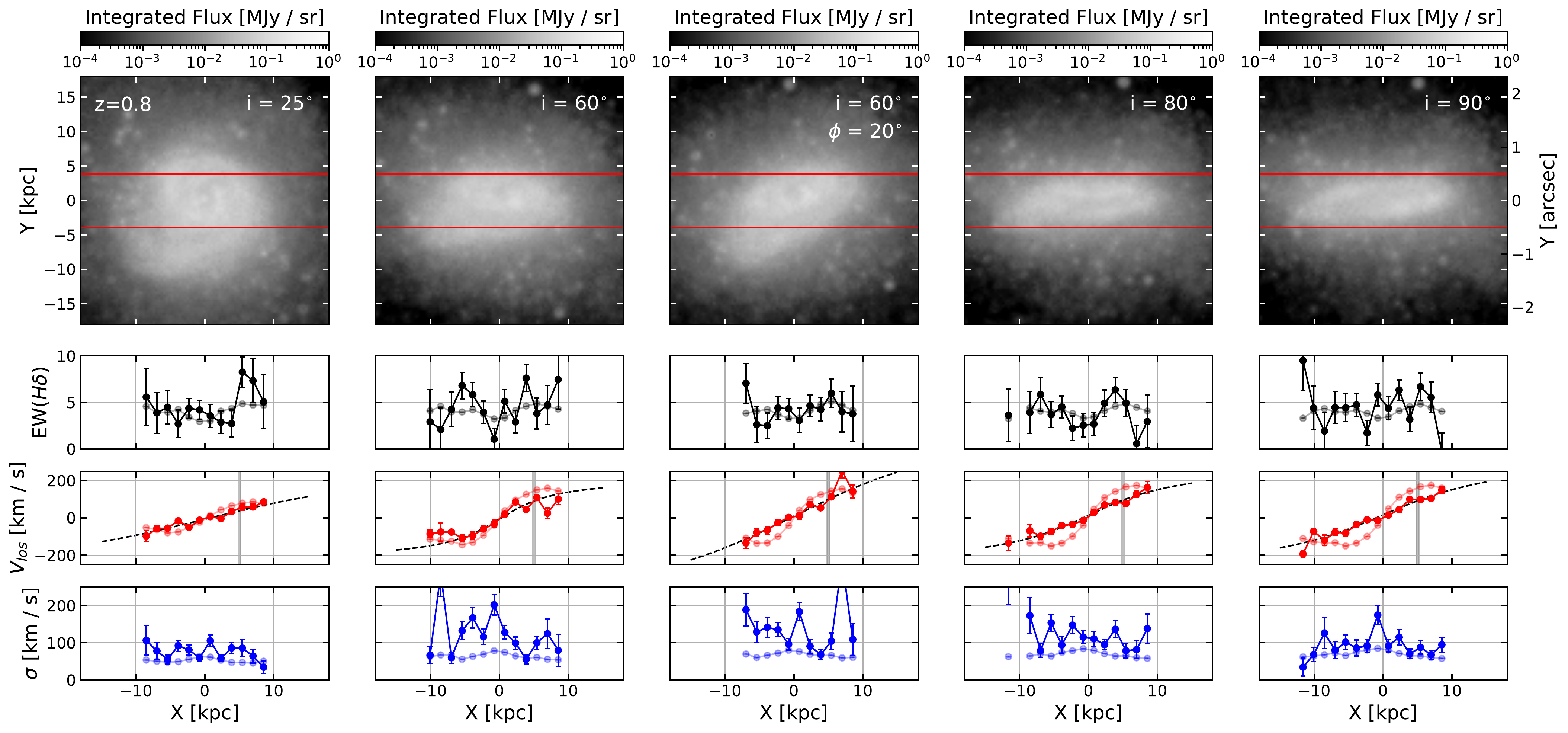}
  \caption{Example of our LEGA-C-like synthetic observations for halo Au-6 in the Auriga sample at $z=0.8$. The rows show from top to bottom, the unattenuated flux distribution (a), $\rm{H\delta}$ equivalent width (b), line-of-sight velocity (c) and velocity dispersion (d) curves. Realisations of the synthetic observation at different angles are shown on each column, increasing with inclination to the right. For each of the curves in the 3 bottom rows, a version in a lighter shade was plotted indicating the intrinsic values, obtained from mock observations without post-processing. Only pixels along the slit with sufficient photon packets in the~\texttt{SKIRT} run are displayed, as the rest are deemed unreliable. The mock-slit position is marked in red at the top panel and on the 3rd row a grey dashed line indicates the best-fit arctangent model to the rotational velocity curve, used to calculate the rotational velocity at 5 kpc ($V_5$).}\label{fig:legac-summ}
\end{center}
\end{figure*}

\subsubsection{Spectral fitting and analysis}\label{sec:spec-fit-legac}

Analogous to the analysis performed for the low $z$ synthetic spectra, we perform a spectral fit with pPXF to the mock 2D LEGA-C observations. We use a procedure similar to the one described in Section~\ref{sec:spec-fit-sami} with two key differences. Since our synthetic observations are generated in the observed frame at high $z$ we first bring them to rest-frame using the redshift of the simulation. Also, instead of using median stacked spectra for annuli at different radii to obtain optimal templates, the first fit we perform is on the 1D integrated spectrum. The resulting set of templates then is used to fit each of the pixels along the slit and get the line-of-sight velocity and velocity dispersion curves. 

Figure~\ref{fig:legac-summ} showcases the resulting line-of-sight rotation velocity and velocity dispersion curves along the slit for the simulated spectra of the Auriga halo Au-6 at the different inclinations. Compared to the intrinsic v$_{\rm{los}}$, derived from the unattenuated spectra, we see shallower slopes in the rotation curves for our post-processed spectra. Both observations reach similar velocities at the outskirts, within the measurement uncertainty, for $r\lesssim10$~kpc, but the  velocities near the centre are higher than those estimated for the attenuated spectra. These offsets become as high as $70$ km s$^{-1}$ around the $r=5$ kpc distance. As mentioned in the previous section, a similar effect had already been reported on by~\citet{Baes2003RadiativeGalaxies}. However, these effects become almost negligible at $i \leq 60\degree$ for their disk models, in contrast with our sample. Beam smearing due to the blending of light would also contribute to the flattening of the profile. In the case of the velocity dispersion profiles, we measure consistently higher dispersions for the attenuated spectra with respect to the stellar-only intrinsic values. This effect can be associated with the PSF limitations, which for the LEGA-C sample is as large as the size of the slit. We also compute the equivalent width of the $\rm{H}\rm{\delta}$ absorption feature across the galaxy's major axis as shown on the second row of Figure~\ref{fig:legac-summ}. This illustrates that recovering subtle variations in stellar population properties at this resolution is challenging (but see~\citet{DEugenio2020InverseLEGA-C}).

To estimate the effects of radiative transfer and the instrumental and observational limitations on the derived rotational support we compare $V/\sigma$ for the different realizations (total, radiative transfer only and instrumentals only) across all inclinations simulated. We fit the rotation curves derived from \texttt{pPXF} with an arctangent model using an MCMC procedure and measure the line-of-sight rotational velocity at 5 kpc, $V_5$, as the value of the best-fit model at this distance from the centre of the slit, as defined in~\citet{Bezanson2018Galaxies}. Uncertainties for these values correspond to the 16th and 84th percentiles in the posterior distribution. We also measure the central velocity dispersion, $\sigma_0$, as the mean value of the velocity dispersion profile of the two pixels at the centre of the slit. As for the SAMI-mocks we provide "intrinsic" values for these estimates by repeating this procedure on the unattenuated synthetic observations at an edge-on inclination. 

Similarly to the Local Universe case, the rotation support recovered as $V/\sigma$ is lower than the estimated for the "ground truth" case across all inclinations for our fiducial run, as illustrated in Figure~\ref{fig:vsigma-all} by the magenta squares. The trend with inclination is less robust as the one observed at $z=0$, as the uncertainty of the measured $V/\sigma$ is larger for this mock observations. In contrast with the low $z$ sample, the offsets observed are a result of the spatial convolution principally. The dust attenuation only produces underestimations on the order of 0.09 dex on the measured $V/\sigma$, at its highest, while instrumental effects, mainly the seeing convolution lead to values 0.5 dex lower than intrinsic. On average these offsets produce an underestimation of $V_5/\sigma_0$ by a factor $\sim 3$. 

\subsection{Simulating HARMONI IFS spectroscopy at z=3}\label{sec:z=3}

\subsubsection{ELT - HARMONI}\label{sec:harmoni}

\begin{figure*}
\begin{center}
     \includegraphics[width = \textwidth]{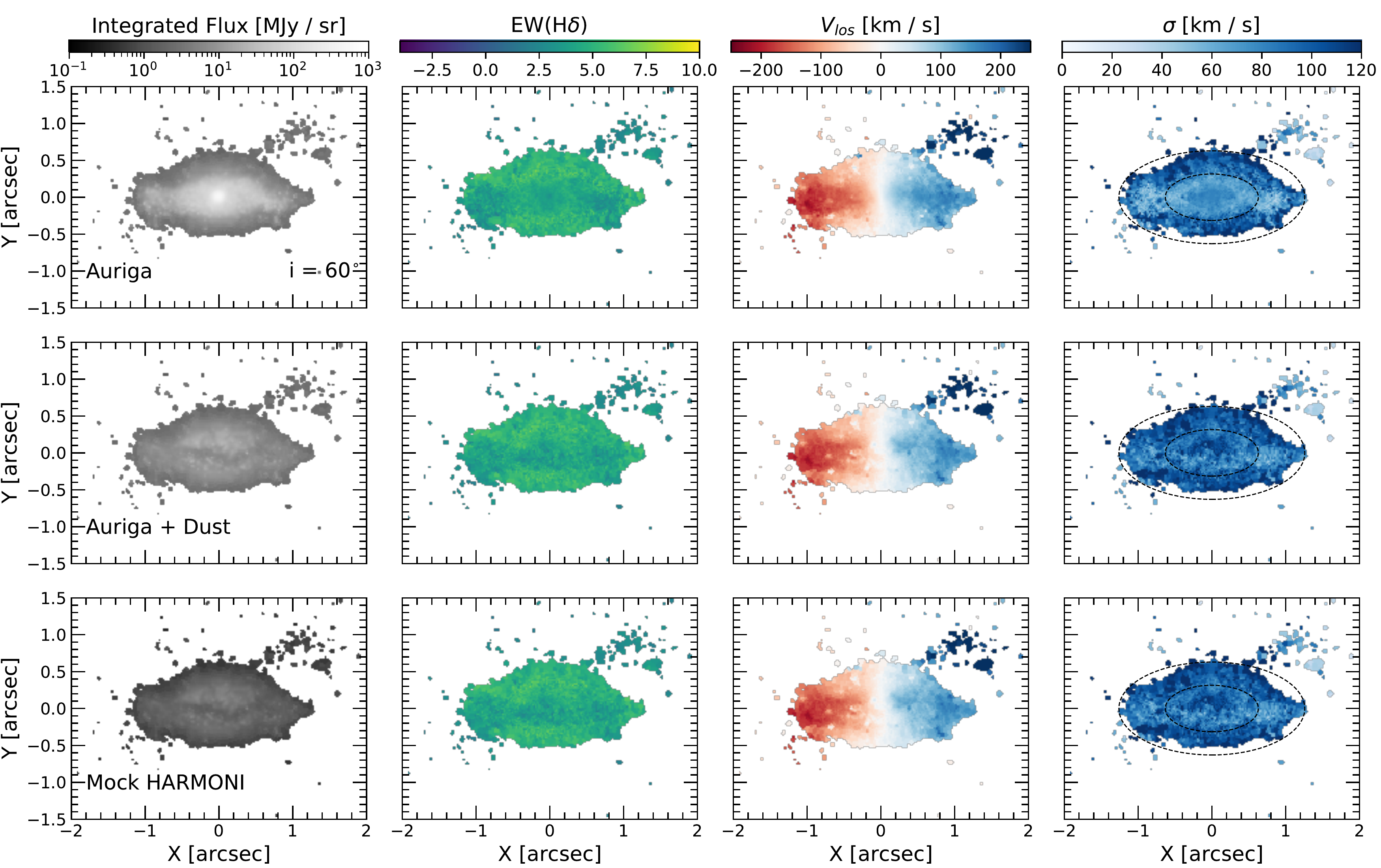}
  \caption{Resulting flux (a), $\rm{H\delta}$ equivalent width (b), line-of-sight velocity (c) and velocity dispersion (d) distributions for our synthetic HARMONI mock observations of halo Au-29 of the Auriga sample at z=3. The inclination of the observation shown is set to $i=60\degree$. The top row shows the resulting distributions from unattenuated unconvolved IFS-like data cubes. In the middle, radiative transfer with~\texttt{SKIRT} is added and on the bottom row the maps are the result of our \texttt{SKIRT} run and further processing i.e. PSF convolution and noise addition. Ellipses on the last column (d) denote 1 $R_{\text{e}}$ and 2 $R_{\text{e}}$.}\label{fig:harmoni_summary}
\end{center}
\end{figure*}

As a final application of our methodology we produced synthetic spectroscopy emulating one of the set-ups of the High Angular Resolution Monolithic Optical and Near-infrared (HARMONI) instrument at the ELT. This integral field spectrograph will be one of the first generation instruments for ELT and is expected to provide observations in a variety of spatial and spectral resolutions within $0.47$--$2.45 \rm{\mu m}$. 

Our experiment's goal is to test the capabilities of HARMONI at observing the rest-frame optical stellar continuum, and its absorption features, for a galaxy at $z\sim 3$. The most appropriate set-up then is to use the H band coverage at a spectral resolution of $R=3500$. The field-of-view chosen is the 3.04" x 4.08" set-up which comes with a pixel resolution of 20 mas on the side. At $z=3$ this configuration spans a 31.62 x 23.56 kpc area, enough to cover at least 1 R$_e$ for the haloes in the Auriga sample. 

Since the $z\sim 3$ galaxies in the Auriga simulations are not sufficiently massive for our purpose, we use a $z\sim 2$ snapshot of Au-29, which has a stellar mass of $M_\star=1.65\times10^{10}$ M$_{\odot}$ at that redshift and a dust mass fraction of $0.0175$. Although the snapshot used for the extraction corresponds to a lower redshift, we run the radiative transfer procedure at a distance corresponding to $z=3$, to obtain the correct rest-frame wavelength coverage of the optical spectral features needed for the kinematic analysis.

\subsubsection{Generation of the HARMONI-like data cubes}\label{sec:harmoni-mock}
Since HARMONI is an IFS instrument, we use a configuration for our \texttt{SKIRT} simulations analogous to the $z=0$ case. We chose a pixel scale of 0.155 kpc on the side and a wavelength range at the observed frame (at $z=3$) from 1.4 to 2.2 $\mu$m, which covers the rest-frame range between 3500 -- 5500 \AA, as can be seen in Fig~\ref{fig:spec-range}. We generated cubes at inclinations $i=0\degree, 60\degree, 80\degree$ and $90\degree$, also allowing for the separate recording of the components to obtain the unattenuated data cubes. 

The further manipulation of the data cubes to resemble observations with HARMONI is analogous to the procedure described in Section~\ref{sec:obs-effects} for the SAMI Galaxy Survey mocks. We start by convolving the slices at each wavelength with a Moffat profile, using parameters $\beta=4.765$ and FWHM=10 mas. The value of the PSF FWHM was chosen to be the driffaction limit of the ELT at this wavelength as we simulate observations assuming a perfect Adaptive Optics (AO) correction. 
To add noise we assume a constant noise across wavelength range, for simplicity, and set the signal-to-noise of our observations to S/N$=5$ at 1 R$_e$. Using this parameters we perturb the resulting fluxes in our data cubes.  

\subsubsection{Spectral fitting and analysis}\label{sec:spec-fit-harmoni}

As with the rest of our mocks we perform kinematic and equivalent width fitting following a procedure analogous to that for our synthetic SAMI observations, as outlined in Section~\ref{sec:spec-fit-sami}. The resulting maps for the intrinsic, stellar only (top), attenuated (middle) and seeing convolved (bottom) mocks are displayed in Figure~\ref{fig:harmoni_summary}. The integrated flux image illustrates the effect of attenuation on our simulated spectra, and we see an evident change to the light distribution. The bright centre of the galaxy, shown in the first row, is obscured by the dust in the line-of-sight.

The HARMONI mock observation, despite strong attenuation, accurately recovers the overall kinematic structure, demonstrating that a dynamical model based on a strongly attenuated spectrum will still produce an accurate description of the the mass distribution and the dynamical structure. Stellar population gradients are also recovered with good precision thanks to the superior spatial resolution, but this has a downside: if the stellar population $M/L$ estimate is insensitive to attenuation, then the true $M/L$ will be difficult to measure. While beyond the scope of this paper, our methodology allows for quantitative tests to address this issue.

\begin{figure*}
\begin{center}
     \includegraphics[width = \textwidth]{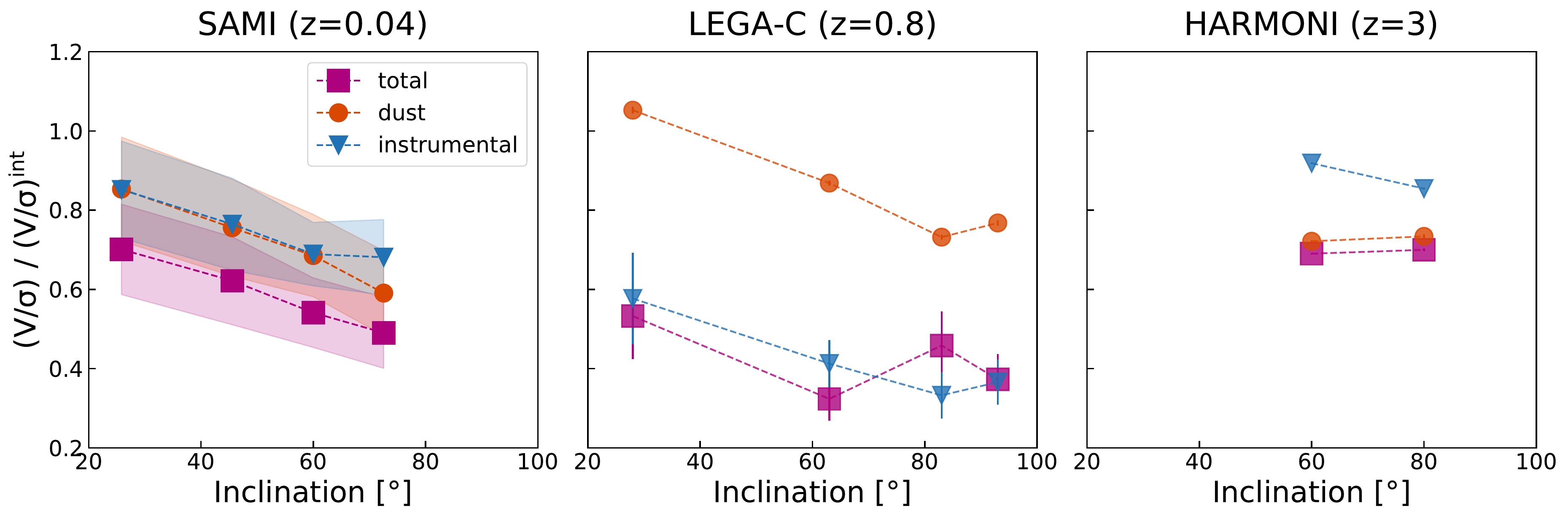}
  \caption{Effect of post-processing on the measured rotational support of the galaxies in the Auriga sample. (Left) Data points correspond to the median offsets in the computed $V/\sigma$ within 1 $R_e$ for our sub-sample of mock SAMI-like IFS cubes at each inclination used. The different symbols and colours represent the processes included in the generation of the data cubes, i.e. radiative transfer (orange circles), PSF convolution and noise addition (blue triangles) and both (magenta squares). The shaded areas correspond to 1 $\sigma$ of the distribution of offset values at each inclination. At this redshift, instrumental and dust effects produce similar offsets, with the effects becoming more prominent at higher inclinations. (Centre) Offsets in the computation of $v_5/\sigma_0$ for halo Au-6 at $z=0.8$. Rotational velocities and velocity dispersion were calculated from synthetic long-slit observations emulating the LEGA-C survey data. Errorbars correspond to the 1$\sigma$ uncertainties in the measurements of the properties. In contrast to the SAMI mocks at a lower $z$, LEGA-C has a lower spatial resolution resulting in instrumental effects being more prominent. (Right) $V/\sigma$ within 1 $R_e$ for the synthetic IFS data generated for halo Au-29 at $z=3$. The data cubes were generated using the specs of future ELT instrument HARMONI. In this case the effects of dust become more prominent than the seeing effects, due to the high resolution of HARMONI and the high amounts of dust of the source selected. At all redshifts the instrinsic values used for the comparison were computed using unattenuated and unperturbed versions of the synthetic observations. Rotation velocities were calculated from an edge-on (i=90\degree) orientation while velocity dispersions used face-on (i=0\degree) views.}\label{fig:vsigma-all}
\end{center}
\end{figure*}

Finally, we recover global kinematics for the synthetic cubes at inclinations $i=60\degree$ and $i=80\degree$. We use the same definitions for $V_{\text{rot}}$ and $\sigma_{\text{e}}$ from Equations~\ref{ec:vrot} and~\ref{ec:sigma}, respectively, and compare with "ground truth" values measured from the unattenuated, unconvolved data cubes. In our fiducial run, using the full procedure including radiative transfer and instrumental effects, we see an underestimation of rotational support in both inclinations simulated, as presented in Figure~\ref{fig:vsigma-all}. The amplitude of the offset, however, is much smaller compared to both lower $z$ experiments, and at its maximum produces an underestimation of 0.15 dex in the $V/\sigma$ measured within 1$R_{\rm{e}}$. This difference might be due to the better spatial resolution of HARMONI. We see that the run that only implements the observational effects produces only very little offsets, lower than 0.1 dex, while most of the underestimation is caused by the presence of dust, in contrast to what we observe in our LEGA-C exercise at $z=0.8$. Due to the reduced dynamical range of the inclinations covered we can't establish any trends between the amplitude of the offsets and inclination.  

\section{Summary and Outlook}\label{sec:summary}

We present a framework for producing synthetic spatially resolved galaxy spectra from cosmological hydrodynamical simulations. With the 3D radiative transfer code~\texttt{SKIRT}~\citep{Baes2020InfraredSimulation} we propagate high-resolution synthetic spectra from~\citet{Conroy2009TheGalaxies} through the dusty interstellar medium of the high-resolution hydrodynamical simulation Auriga~\citep{Grand2017TheTime}. The RT simulation is converted into mock integral field or slit spectroscopic observations, including instrument specifications, noise and the point-spread function.  
As an application of our method we show mock observations of the Milky Way mass Auriga simulations that match the instrumental specifications of the SAMI Galaxy survey of present-day galaxies, the VLT/VIMOS LEGA-C survey at $z\sim 0.8$, and, looking ahead, E-ELT/HARMONI observations at $z=3$. Stellar kinematic measurements (velocities and velocity dispersions) are performed in the same manner as real observational data. 

A comparison between 10 galaxies in the Auriga sample and the SAMI Galaxy survey galaxies shows that the kinematic signatures of these simulated galaxies show a good agreement with the observed scaling relations after introducing the dust treatment and instrumental effects. The intrinsic velocity dispersion for the simulated Auriga galaxies is systematically offset from the observed Faber-Jackson relation with our velocity dispersion measurements falling below all galaxies in the SAMI Galaxy Survey at a similar mass range. From our analysis we conclude that the discrepancy in simulated and observed kinematics is due to both the effects of dust attenuation and the limited spatial resolution in the observed sample. 

Comparisons with unattenuated and unperturbed spectra show that in general our mocks produce lower line-of-sight velocities and higher velocity dispersions than the intrinsic ground-truth values. These offsets are larger for highly inclined discs, an effect that can be attributed to the spatial convolution of the individual spaxels but is also dependent on the optical depth of the interstellar medium. Which of these effects is most prominent depends on the amounts of dust in the system and the quality of the data, specifically the spatial resolution. This is highlighted in Figure~\ref{fig:vsigma-all}, where the varying spatial resolutions of the surveys/instruments emulated leads to different levels of underestimation. The combination of these offsets can cause an underestimation of V/$\sigma$ by a factor $\sim 2$ -- 3 in our local and high $z$ mocks. 

In summary, our work serves three broad purposes:

\begin{itemize}
    \item The ability to produce data cubes of synthetic spectra that can be treated as observations provides an opportunity to test the techniques and assumptions used to derive physical quantities from observed data. The comparison with the intrinsic values from the input simulations allows for the study of the biases introduced by the underlying assumptions and inherent imperfections due to finite $S/N$ and spatial resolution.

    \item Our method offers a robust way to compare simulations and observations, that take into account the effects of dust, seeing, noise and the biases introduced by the models used to infer properties from observations. These ultimately inform the physical models and recipes used to calibrate the sub-grid recipes the simulations rely on. 
    
    \item With our framework that combines simulated galaxies and radiative transfer we can provide more realistic and physically motivated input models to test the capabilities of future instrumentation.  
\end{itemize}

This work focused on the stellar kinematics derived from spatially resolved data, but our methodology can be used to asses the impact of instrumental effects and dust attenuation on other physical properties derived from spectra, such as stellar population parameters (e.g. ages, metallicities, etc.). We plan to build on this work by including a physical treatment of the radiative transfer of line emission (Kapoor et al., in prep.). This will enable similar studies of the gas kinematics and the inferred properties of the ISM derived from emission line strengths and shapes and their biases. And will also provide a robust method of comparison of the simulated galaxies with the rich available datasets of spatially resolved H$\alpha$ kinematic maps from SINFONI~\citep{Tacchella2015TheImaging} and KMOS~\citep{Tiley2021TheGyr}.

\section*{Acknowledgements}
We thank Luca Cortese and Eric Emsellem for useful comments and suggestions. This project received funding from the European Research Council (ERC) under the European Union’s Horizon 2020 Framework Programme, grant agreement no. 683184 (Consolidator Grant LEGA-C). D.B, M.B and A.U.K acknowledge the financial support of the Flemish Fund for Scientific Research (FWO-Vlaanderen), research projects G039216N and G030319N.

R.G acknowledges financial support from the Spanish Ministry of Science and Innovation (MICINN) through the Spanish State Research Agency, under the Severo Ochoa Program 2020-2023 (CEX2019-000920-S) and an STFC Ernest Rutherford Fellowship (ST/W003643/1).

\section*{Data Availability}
The resulting IFS-like data cubes produced with~\texttt{SKIRT} for the 10 galaxies at z=0 and the two galaxies at z=0.8 and z=3 respectively will be publicly available at \url{www.auriga.ugent.be}
The scripts used to manipulate and add the instrumentals to the original SKIRT output data cubes can be found in \url{https://github.com/dbarrientosa/synth-spec}.


\bibliographystyle{mnras}
\bibliography{references} 




\appendix



\bsp	
\label{lastpage}
\end{document}